\documentclass[
 reprint,
 amsmath,amssymb,
 aps,
]{revtex4-1}

\usepackage{natbib}

\usepackage{graphicx}
\usepackage{dcolumn}
\usepackage{bm}

\newcommand{\ket}[1]{ | #1 \rangle}

\begin{document}

\preprint{APS/123-QED}

\title{Creation of a low-entropy quantum gas of polar molecules in an optical lattice}

\author{Steven A. Moses}
\affiliation{JILA, National Institute of Standards and Technology and the University of Colorado, and the Department of Physics, University of Colorado, Boulder, CO 80309, USA}
\author{Jacob P. Covey}
\affiliation{JILA, National Institute of Standards and Technology and the University of Colorado, and the Department of Physics, University of Colorado, Boulder, CO 80309, USA}
\author{Matthew T. Miecnikowski}
\affiliation{JILA, National Institute of Standards and Technology and the University of Colorado, and the Department of Physics, University of Colorado, Boulder, CO 80309, USA}
\author{Bo Yan}
\altaffiliation[Current Address: ]{Department of Physics, Zhejiang University, Hangzhou, China 310027.}
\author{Bryce Gadway}
\altaffiliation[Current Address: ]{Department of Physics, University of Illinois at Urbana-Champaign, Urbana, IL 61801, USA.}
\author{Jun Ye$^{\ddagger}$}
\affiliation{JILA, National Institute of Standards and Technology and the University of Colorado, and the Department of Physics, University of Colorado, Boulder, CO 80309, USA}
\author{Deborah S. Jin}\thanks{To whom correspondence should be addressed: ye@jila.colorado.edu, jin@jilau1.colorado.edu}
\affiliation{JILA, National Institute of Standards and Technology and the University of Colorado, and the Department of Physics, University of Colorado, Boulder, CO 80309, USA}

\date{\today}

\begin{abstract}
Ultracold polar molecules, with their long-range electric dipolar interactions, offer a unique platform for studying correlated quantum many-body phenomena such as quantum magnetism.  However, realizing a highly degenerate quantum gas of molecules with a low entropy per particle has been an outstanding experimental challenge.  In this paper, we report the synthesis of a low entropy molecular quantum gas by creating molecules at individual sites of a three-dimensional optical lattice that is initially loaded from a low entropy mixture of K and Rb quantum gases. We make use of the quantum statistics and interactions of the initial atom gases to load into the optical lattice, simultaneously and with good spatial overlap, a Mott insulator of bosonic Rb atoms and a single-band insulator of fermionic K atoms. Then, using magneto-association and optical state transfer, we efficiently produce ground-state molecules in the lattice at those sites that contained one Rb and one K atom. The achieved filling fraction of 25\% indicates an entropy as low as $2.2~k_B$ per molecule. This low-entropy molecular quantum gas opens the door to novel studies of transport and entanglement propagation in a many-body system with long-range dipolar interactions.
\end{abstract}

\maketitle

Polar molecules are an ideal candidate system for studying spin physics and emulating quantum magnetism~\cite{toolbox, demlermagnetism, qmagreview, rbcs}.  However, low temperatures and long lifetimes are required.  Ultracold fermionic KRb molecules have been created at a temperature, $T$, close to the Fermi temperature, $T_F$~\cite{kangkuen}, but cooling the trapped gas deeply into quantum degeneracy has yet to be demonstrated.  The largest obstacle arises from the fact that two KRb molecules can undergo a chemical reaction and this limits the lifetime of the trapped gas~\cite{chemistry}. Furthermore, the chemical reaction rate increases in an applied electric field because of the attractive part of the dipole-dipole interactions~\cite{chemistry}.  A solution to this problem is to confine the molecules in a deep optical lattice in order to restrict collisions~\cite{buchlerpotential, marcio, amodsenprl}.  In particular, the lifetime of ground-state molecules in a deep three-dimensional (3D) lattice was demonstrated to be longer than 20 s and limited by off-resonant scattering of the lattice light~\cite{amodsenprl}.  With the chemical reactions mitigated, the remaining challenge is to create a low entropy system, which in the lattice corresponds to  increasing the filling fraction.  In this paper we report the realization of a high-filling, low-entropy quantum gas of ground-state molecules in a deep 3D lattice using a quantum synthesis approach.  

Simulating quantum many-body physics with lattice-confined atoms requires a filling near unity and correspondingly low entropy~\cite{qmagatomsreview}. This condition can be significantly relaxed with polar molecules thanks to their long-range dipolar interactions, which allow for a decoupling of spin (rotational state of the molecules) and motion so that only the spin entropy, which can be prepared to be near zero, is relevant~\cite{kadenproposal}.  This was recently demonstrated in Refs.~\cite{bonature, mbdynamics}, where a spin-1/2 system was realized by encoding spin in the rotational degree of freedom of the KRb molecules.  At dilute lattice fillings, spin exchange via dipolar interactions was observed in the density-dependent decay of, and oscillations in, the spin coherence.  In order to go beyond the observation of dipolar spin-exchange interactions and explore new scientific frontiers, such as studying the spin-$1/2$ Hamiltonian for quantum magnetism~\cite{tjvw, crmagnetism, munichspin, monroe, blatt, diamond, rydberg}, the propagation of excitations and the growth of entanglement and correlations \cite{correlations, johannesent}, many-body localization~\cite{mblocalization}, exotic quantum phases~\cite{dwave, wigner, salvatoretopology, supersolid, yaochern}, and spin-orbit coupling with molecules~\cite{socoupling}, higher lattice fillings will be essential.  Determining what constitutes high lattice filling depends on the specific experiment in question; however, for dynamics studies, one benchmark is the percolation threshold, which for an infinite simple cubic lattice with nearest neighbor interactions corresponds to a filling $\sim 0.3$~\cite{percolation}.  Because of the molecules' long-range interactions and the finite system size, a filling near this percolation threshold is sufficient for exploring dynamics such as the propagation of excitations.

The original success in realizing a nearly quantum degenerate gas of polar molecules~\cite{kangkuen} relied on devising techniques to make ground-state molecules from an ultracold atom gas rather than directly cooling the molecules.  Continuing in this general approach and to sidestep the difficulty in direct cooling of molecules, our strategy for realizing higher lattice fillings for polar molecules is to take advantage of the precise experimental control available for manipulating the initial atomic quantum gas mixture in a 3D lattice. While this basic approach has been proposed in a number of papers~\cite{zollerprop, freericks}, it is very challenging to realize experimentally.  Specifically, one needs to prepare a low entropy state of two atomic species in the lattice and combine this with efficient molecule production. Our molecule production uses magneto-association to first create very weakly bound Feshbach molecules followed by optical transfer to the molecular ro-vibrational ground state. In previous work, we showed that the conversion efficiency from atoms to Feshbach molecules is high ($87\pm13\%$) for lattice sites containing exactly one Rb atom and one K atom~\cite{amodsenprl}. In addition, previous measurements of inelastic collisional loss rates for Feshbach molecules with K or Rb atoms~\cite{chemistry, zirbel} suggest that having an extra atom on a lattice site will be detrimental to molecule production at that site.

 \begin{figure}
\includegraphics[width=8.8cm]{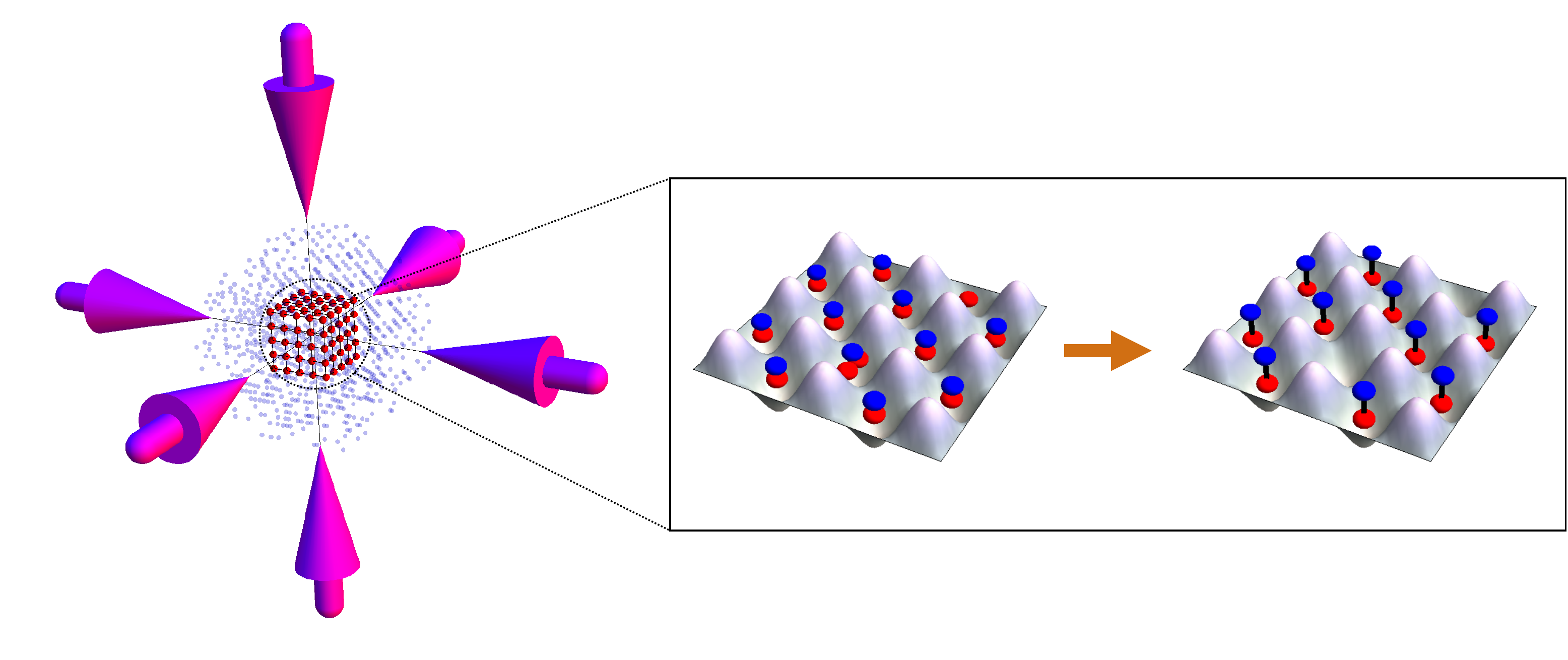}
\caption{\label{fig:figmotivation}Quantum synthesis for creating polar molecules.  Left: strategy for realizing high filling of molecules in a 3D lattice.  We load K (blue) and Rb (red) atoms into a 3D optical lattice, with many more K atoms than Rb atoms.  In the center of the lattice where the two atom clouds overlap, we realize a Rb Mott insulator and a K single-band insulator, each with near unity filling. Right: Zoom-in showing molecule production.  Sites with one Rb and one K have a high probability of producing molecules, while sites with multiple Rb or with only a single atomic species do not yield molecules.}
\end{figure}

The basic scheme is illustrated in Fig.~\ref{fig:figmotivation}. By loading a nearly pure Bose-Einstein condensate (BEC) of Rb atoms into a 3D optical lattice, we can achieve a Mott insulator (MI) state. Here, repulsive interactions between the Rb atoms drive a transition to a state that has an integer number of particles per site~\cite{latticereview}, and
the lattice depth is subsequently increased to pin the Rb atoms. For making molecules, the initial BEC density should be sufficiently low to avoid having multiply occupied sites. For spin-polarized fermionic K atoms, Pauli blocking will prevent any site from having more than one K atom if the atoms are all prepared in the lowest band.  The optimum case is a K band insulator \cite{fermibandinsulator,fermimottinsulator} of one atom per site, which requires starting with a relatively large initial K density.
While a MI of Rb and a band insulator of K are relatively straightforward to achieve separately, creating both simultaneously is very challenging.  The densities of both the Rb and K gases should be $\sim (\lambda/2)^{-3}$ prior to loading the lattice, where $\lambda/2$ is the lattice spacing.  When loading both species into a common optical lattice, we thus need to work with a Rb BEC with small atom number and a degenerate Fermi gas with large atom number. The Rb MI must be well spatially overlapped with the center of the much larger K distribution.  We also need to preserve the high filling of each atomic species in the presence of the other. For this, control over the interspecies interactions is an essential tool. Finally, any excess atoms should be removed from the lattice after the molecule production. 

To prepare the atomic quantum gases, we evaporate Rb in the $\ket{1,1}$ state and sympathetically cool K in the $\ket{9/2,-9/2}$ state in a crossed-beam optical dipole trap with a wavelength $\lambda =1064$~nm.  Here, the atomic hyperfine states are denoted by $\ket{F,m_F}$, where $F$ is the total atomic spin and $m_F$ is its projection. The evaporation is performed at a magnetic field, $B$, of 540 Gauss, where the interspecies scattering length, $a$, is $-100~a_0$, where $a_0$ is the Bohr radius.  This field provides for modest interactions between the two atomic species while being close to an interspecies Feshbach resonance~\cite{fbrparameters} at $B_0=546.6$~G that is used for tuning of the interactions as well as for the molecule creation.  The final optical trap is cylindrically symmetric with a typical axial trap frequency of $\omega_z = 2 \pi \times 180$ Hz (in the vertical direction) and a radial trap frequency of $\omega_r=2 \pi \times25$ Hz for Rb. The measured trap frequencies for K are $2 \pi \times 260$ Hz and $2 \pi \times 30$ Hz.  The larger vertical trap frequency helps prevent separation of the Rb and K clouds due to gravitational sag. Immediately after the evaporation, we turn off the interspecies interactions by ramping $B$ to 543.6 G where $a=0$.  At this point, we have a  Fermi gas of between $1\times 10^5$ and $2 \times 10^5$ K atoms and a nearly pure Rb BEC with $10^3$ to $10^4$ atoms.  Once the Rb BEC forms, Rb no longer thermalizes efficiently with K, and as a result the temperature of the K gas is limited to $T/T_F \approx 0.3$.  We then smoothly turn on, in 150 ms, three retro-reflected beams with $\lambda=1064$~nm that form a cubic optical lattice.   Two of the lattice beams are in the horizontal plane, while the third beam is at an angle of 6$^{\circ}$ from vertical. The final lattice depth is between 20 and 25 $E_{\text{R}}^{\text{Rb}}$, where $E_{\text{R}}^{\text{Rb}}=\frac{\hbar^2k^2}{2m}$ is the recoil energy for Rb, $k=\frac{2\pi}{\lambda}$, and $m$ is the mass of the Rb atom.

\begin{figure*}
\includegraphics[width=14cm]{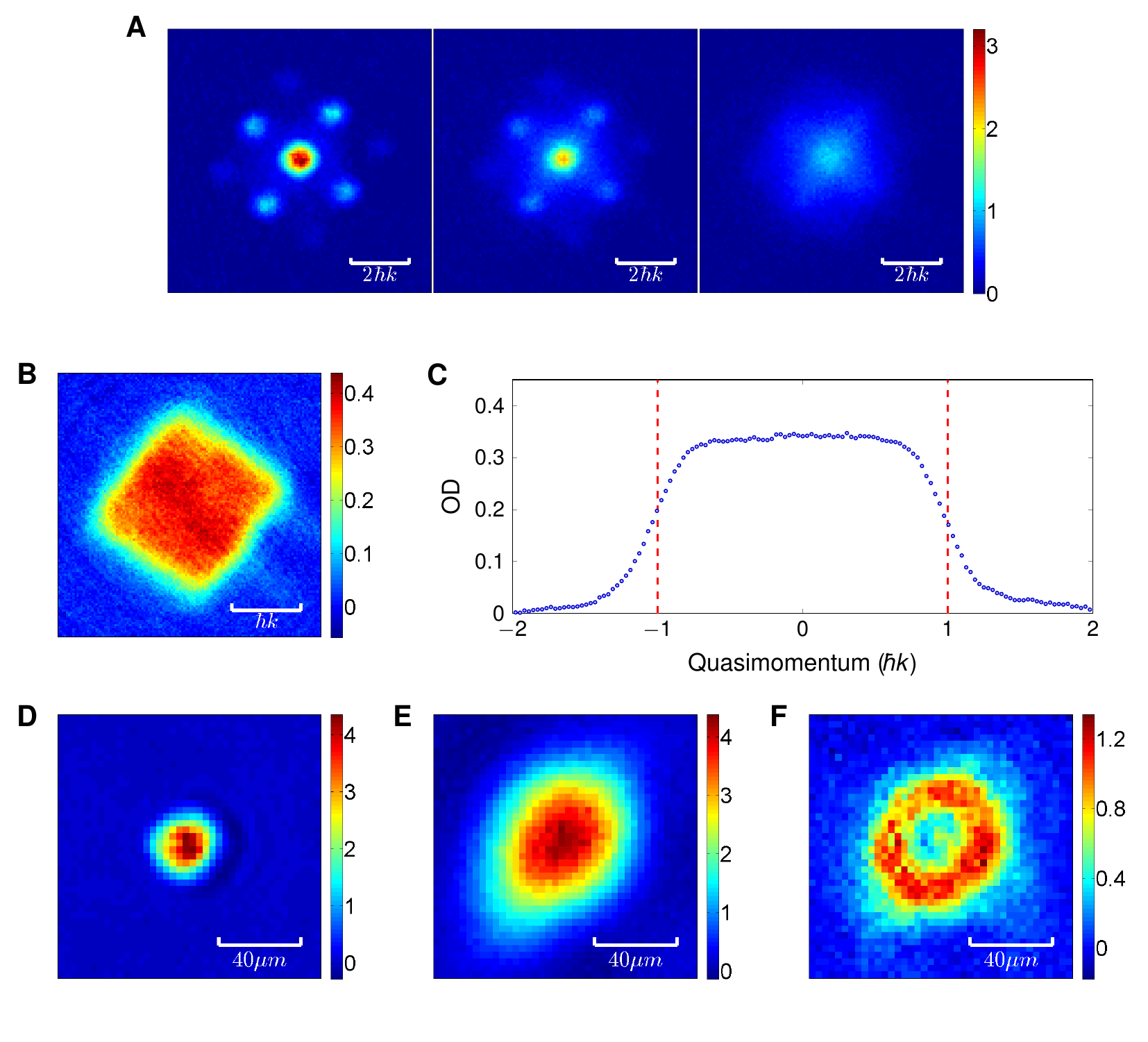}
\caption{\label{fig:pictures}$\textbf{(A)}$ The superfluid-Mott insulator transition for Rb. The three images of Rb were taken after 8 ms of expansion from the lattice, where the final lattice depth is 12, 17, and 22 $E_{\text{R}}^{\text{Rb}}$ from left to right. The optical depth (OD) for each image is indicated by the colorbar to the right of the image.
 $\textbf{(B)}$ Band-mapping of K, imaged after 11.5 ms of expansion. $\textbf{(C)}$ Cut through the K band-mapping image showing the OD vs.~quasimomentum.  $\textbf{(D)}$ $\textit{In situ}$ image of $2\times10^4$ Rb atoms.  $\textbf{(E)}$ $\textit{In situ}$ image of $1.8 \times 10^5$ K atoms.   $\textbf{(F)}$ $\textit{In situ}$ image of the K cloud after initiating loss due to K-Rb inelastic collisions.  The resulting hole in the K cloud demonstrates that the initial spatial overlap with the Rb cloud was good in all three directions.}
\end{figure*}

We image the atom clouds, either $\textit{in situ}$ in the lattice or after a time-of-flight (TOF) expansion, using resonant absorption imaging with a probe beam that propagates along the vertical direction.  Figure \ref{fig:pictures}A shows an example of TOF images of the Rb gas that show the disappearance of coherent matter wave interference as the lattice depth is increased beyond the superfluid-Mott insulator transition. For these images, the number of Rb atoms is $8\times 10^4$ and the final lattice depth in units of $E_{\text{R}}^{\text{Rb}}$ is 12, 17, and 22, from left to right.  Figure \ref{fig:pictures}B shows an image of $1.8 \times 10^5$ K atoms after expansion from the lattice, where the lattice was turned off more slowly for band-mapping \cite{zurichhigherbands}.  A trace through this image along the direction of one of the horizontal lattice beams, which is rotated by roughly $45^{\circ}$ with respect to the camera axes, is shown in Fig.~\ref{fig:pictures}C. This trace, which is averaged along the other horizontal lattice direction, shows that most of the atoms are in the lowest band. The spatial coordinate for the expanded gas image has been converted to quasimomentum in units of $\hbar k$.

Figures \ref{fig:pictures}D and \ref{fig:pictures}E show $\textit{in situ}$ images of Rb and K, respectively.  Note that the Rb cloud is significantly smaller than the K cloud.  To verify that the clouds are spatially overlapped, we use an RF pulse to transfer the Rb atoms to the $\ket{2,2}$ state in order to induce spin-changing collisions that result in loss of K and Rb atoms on the same lattice site.  The resultant hole in the K distribution (Fig.~\ref{fig:pictures}F) clearly demonstrates that the clouds are overlapped in the trap.

We determine the peak filling fraction from fits to the measured atomic distributions.  The K Fermi gas is described by a Fermi-Dirac distribution, which can be approximated by a Gaussian.  In this case the peak filling is:
\begin{equation}
 f_{\text{Gauss}}=\frac{N (\lambda/2)^3}{(2 \pi)^{3/2} \sigma_{x} \sigma_{y} \sigma_{z}},
 \end{equation}
 where $N$ is the number of atoms and $\sigma_x$, $\sigma_y$, and $\sigma_z$ are the Gaussian rms widths.  For the Rb MI, the distribution is better described by a Thomas-Fermi (TF) distribution \cite{rbcalc}.  In this case the peak filling is:
  \begin{equation}
 f_{\text{TF}}=\frac{15 N (\lambda/2)^3}{8 \pi R_{x} R_{y} R_{z},}.
 \end{equation}
 where $R_x$, $R_y$, and $R_z$ are the Thomas-Fermi radii.  We image the gas along $z$, so we determine the radial size.  The vertical size is smaller by a factor of $A =6.4(1)$, which is measured for a thermal gas of Rb in the combined potential of the optical trap and lattice.  
 
 \begin{figure*}
\includegraphics[width=14cm]{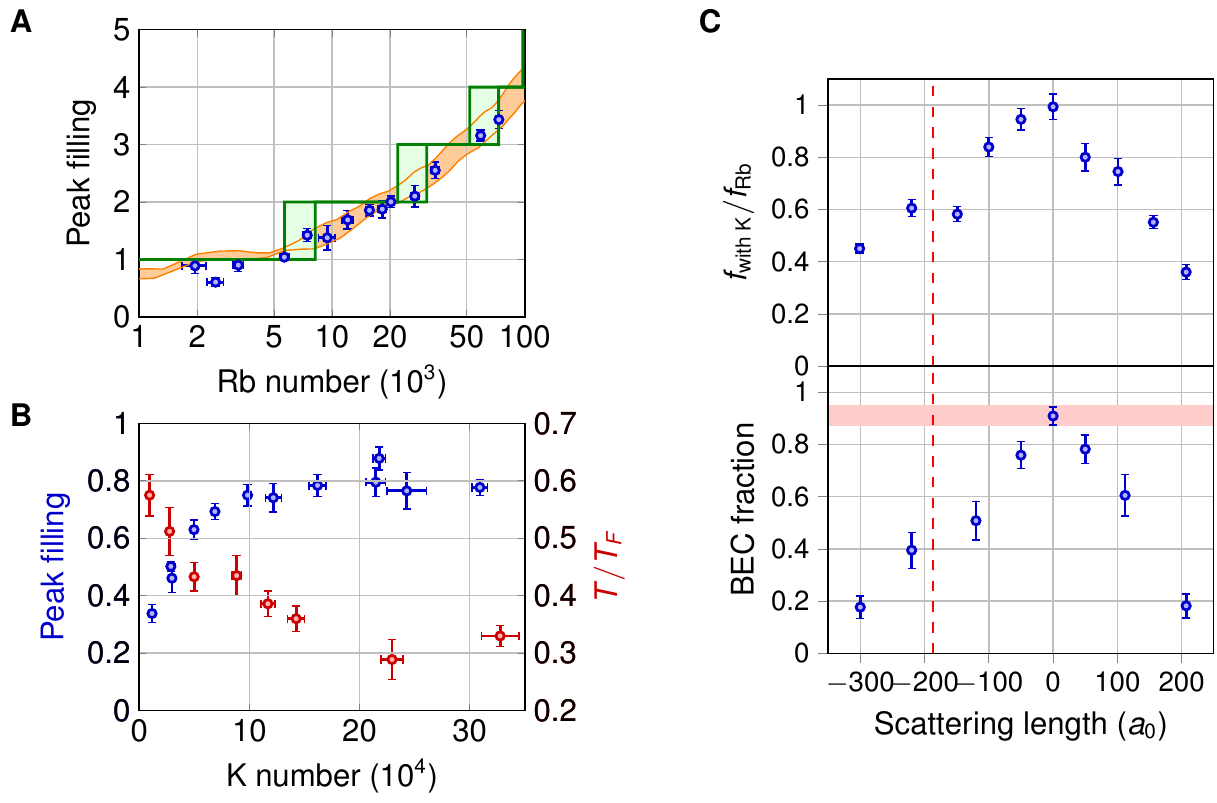}
\caption{\label{fig:figinteraction} $\textbf{(A)}$ Measured Rb MI peak filling $f_{\text{TF}}$ (blue points) vs.~Rb number.  The cloud size was extracted from a TF fit to the \textit{in situ} image, while the number was either extracted from the \textit{in situ} image or a Gaussian fit to the cloud after a few ms of free expansion.  In either case the filling was computed according to Eq. (2) in the text.  The green staircase displays the calculated peak occupancy for a $T=0$ distribution \cite{som}. The total harmonic confinement (including the lattice light) is represented by $\omega_r=2 \pi \times (38 \pm 2)$ Hz and $\omega_z = (6.4 \pm 0.1) \, \omega_r$. The orange band shows a fit to the calculated density distribution, accounting for finite imaging resolution and pixelation present in the experiment. $\textbf{(B)}$ Peak filling $f_{\text{Gauss}}$ of K for a lattice depth of 9$E_{\text{R}}^{\text{K}}$ (blue points), indicating the onset of a K band insulator for $N_{\text{K}} > 10^5$. The red points show the measured $T/T_F$ of the initial K gas before loading the lattice. $\textbf{(C)}$ Top: peak filling of Rb in the lattice in the presence of $1.4 \times 10^5$ K as a function of the interspecies scattering length, normalized to the filling of Rb without K.  Bottom: the initial BEC fraction in the optical trap under the same condition.  Here, the number of Rb atoms is between $2.8 \times 10^3$ and $5.2 \times 10^3$.  The background (non-resonant) scattering length is indicated by the dashed red line.  The red shaded bar indicates the BEC fraction of Rb without K.}
\end{figure*}
 
Figure \ref{fig:figinteraction}A shows the measured peak filling ($f_{\text{TF}}$) for Rb. For a comparison to the data, we calculate the $T=0$ MI distribution for our trap, convolve this distribution with a Gaussian filter to account for the finite imaging resolution, bin the data into pixels, and then fit with a TF distribution~\cite{som}.  For very small samples, the size of the cloud is about twice the imaging resolution.  The data matches well with the $T=0$ calculation, and from this comparison, we infer that the $N=1$ Rb MI occurs for less than 5000 atoms in our trap.

Figure~\ref{fig:figinteraction}B shows the measured peak filling of K ($f_{\text{Gauss}}$) in the lattice. Here, the lattice depth is $23 E_{\text{R}}^{\text{Rb}}$; however, given the different mass and ac polarizability for K, this is equivalent to only $9 E_{\text{R}}^{\text{K}}$, where $E_{\text{R}}^{\text{K}}$ is the recoil energy for K atoms.  We find that the measured $f_{\text{Gauss}}$ rises with increasing K atom number (blue points), and saturates around 80\% for K numbers $\geq$$1 \times 10^5$. For this data, $T/T_F$ decreases with increasing K number (red points). The saturation of the lattice peak filling is consistent with the onset of a band insulator in the center of the lattice.  

The data in Figs.~\ref{fig:figinteraction}A and \ref{fig:figinteraction}B show that to achieve optimal molecule production, the initial BEC should have less than 5000 atoms for a MI with mostly one atom per site, while the Fermi gas should have more than $10^5$ atoms to reach the band insulating limit. When loading both atom species simultaneously, K can affect the filling of Rb, and we find empirically that turning off interactions by going to $a=0$ is optimum. To illustrate this effect, Fig.~\ref{fig:figinteraction}C shows both the measured peak filling of Rb in the lattice and the initial BEC fraction in the optical trap in the presence of $1.5\times10^5$ K atoms as a function of $a$ at the end of the evaporation. We observe a clear dependence, with the highest Rb filling and BEC fraction achieved near $a=0$.  Furthermore, we have checked that the Rb filling at $a=0$ is unaffected by the K for Rb numbers between $2 \times 10^3$ and $10^5$. While the interactions between K and Rb atoms during the lattice loading could affect the MI \cite{freericks, dualybinsulator}, in our data the dominant effect appears to be a higher initial Rb temperature (lower BEC fraction), which results in a poor MI. This higher initial temperature comes from the intrinsic difficulty of sympathetically cooling a large K gas through thermal contact with a smaller number of Rb atoms.
 
\begin{figure*}
\includegraphics[width=14cm]{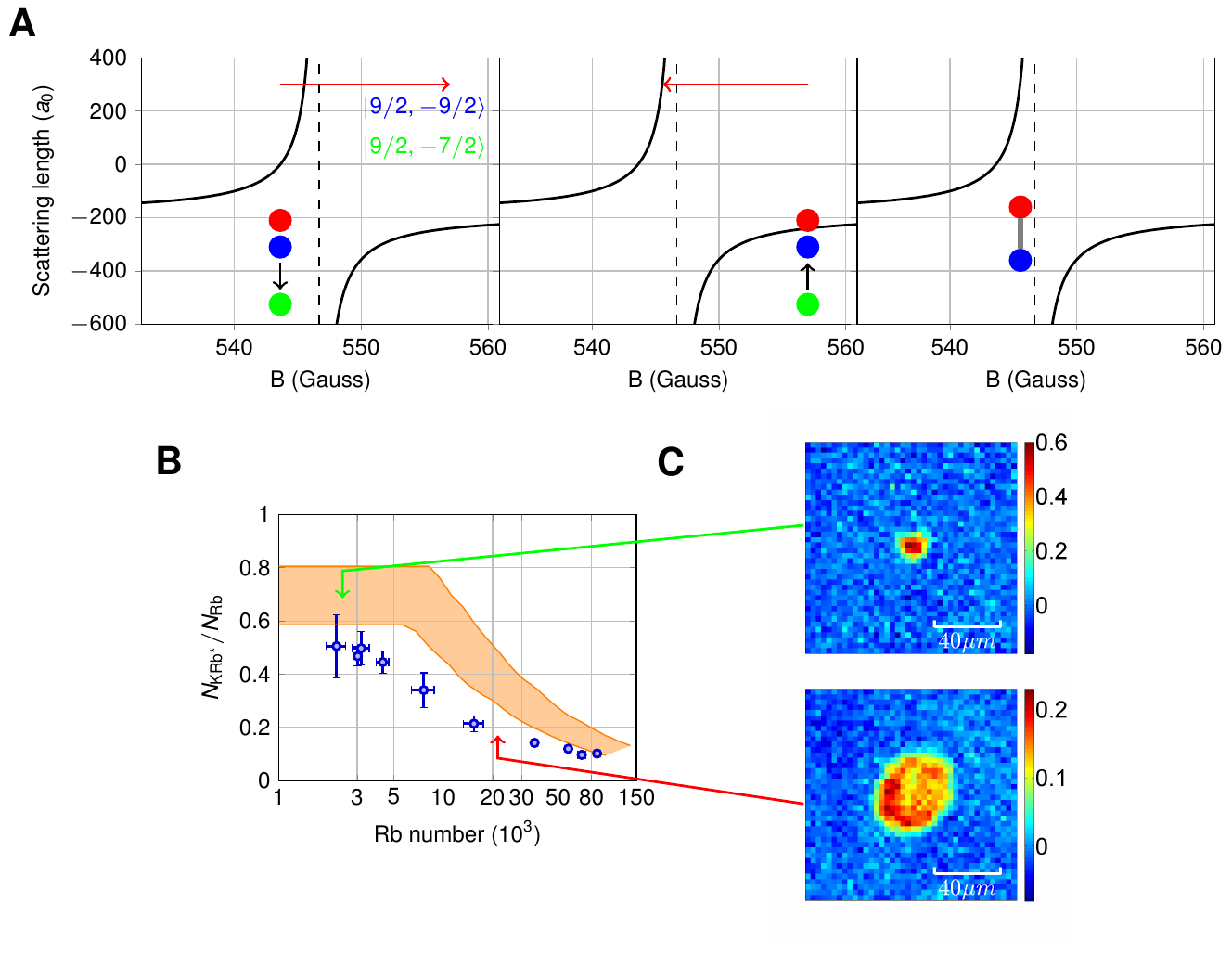}
\caption{\label{fig:figfinal} $\textbf{(A)}$ The interspecies Feshbach resonance. The atoms are loaded into the lattice at $a=0$. The K atoms are then transferred to a hyperfine state ($\ket{9/2,-7/2}$) that does not participate in the resonance, and $B$ is swept above the resonance. After transferring K back to the $\ket{9/2,-9/2}$ state, magneto-association proceeds by sweeping $B$ from above to below the resonance (middle and right panels).  The Feshbach molecules are then transferred to the absolute ground state using STIRAP.  $\textbf{(B)}$ Fraction of Rb atoms converted to Feshbach molecules. The orange shaded region shows the expected fraction of Rb that are on a site with exactly one Rb atom and one K atom. $\textbf{(C)}$ $\textit{In situ}$ images of ground-state KRb molecules in the lattice. The ground-state molecules are held in the lattice for 40 ms before imaging. For low initial Rb number (top image, average of three repeated experiments), we find a filling fraction of $25(4)\%$.  For higher Rb number (bottom image, average of seven shots), we observe a hole in the center of the molecular distribution, and the filling is much less.}
\end{figure*}

In preparing the gas for molecule creation, we find an additional issue arising from interspecies interactions. The first step in creating ground-state molecules is magneto-association, which consists of adiabatically sweeping $B$ across the K-Rb Feshbach resonance from high to low field.  A schematic of the Feshbach resonance and molecule creation process is shown in Fig.~\ref{fig:figfinal}A.  However, starting from $a=0$, which occurs for $B<B_0$, we first need to jump $B$ to the high-field side of the resonance. This jump should ideally be diabatic in order to avoid promoting atoms to higher lattice bands \cite{zurichhigherbands}; however, with the relatively high local atom densities in the lattice, it is difficult to sweep the field fast enough. To overcome this problem, we use an RF pulse to transfer the K atoms to a spin state, $\ket{9/2,-7/2}$, that does not experience the 546.6 G resonance. After ramping $B$ above $B_0$, we transfer the K atoms back to the $\ket{9/2,-9/2}$ state and then proceed with the Feshbach association process.  We find that applying these RF transitions improves the final filling of ground-state molecules by 60\% compared to the case of not doing these RF transitions.

Figure \ref{fig:figfinal}B shows the measured fraction of the Rb number, $N_{\text{Rb}}$, that is converted to Feshbach molecules (blue circles). Since we operate with many more K atoms than Rb atoms, a large background of K atoms remains in the lattice after making molecules. This presents a challenge for determining the number of Feshbach molecules, $N_{\text{KRb*}}$, which we typically measure by dissociating the molecules and imaging K. To selectively count only the molecules, we use an RF pulse to transfer the background K atoms to the $\ket{9/2,-7/2}$ state before dissociating the molecules by ramping $B$ back above the resonance and selectively imaging the K atoms in the $\ket{9/2,-9/2}$ state~\cite{som}.  For comparison with the data, the shaded band in Fig.~\ref{fig:figfinal}B shows the product of the measured $f_{\text{Gauss}}=0.80(5)$ for K, the calculated fraction of Rb atoms of a $T=0$ MI that are on singly occupied sites, and the conversion efficiency of preformed pairs reported in Ref.~\cite{amodsenprl}.  We find that the trend of the calculation matches the data, with the conversion efficiency decreasing for higher Rb number. This is consistent with the assumption that molecules are not produced on sites that have more than one Rb atom.  The data lie slightly below the calculation; possible explanations for this include finite temperature effects on the MI (which could lead to fewer singly occupied sites) or reduced conversion efficiency for sites with one Rb and one K atom. For small Rb numbers, we find that $N_{\text{KRb*}}/N_{\text{Rb}}$ is larger than 50\%.

As a last step, we use stimulated Raman adiabatic passage (STIRAP) to transfer the Feshbach molecules to their ro-vibrational ground state~\cite{kangkuen}. The typical efficiency of this transfer is $89(4)$\%. Once the molecules are in the ground state, we apply resonant light pulses to remove all unpaired atoms from the lattice. This atom removal is essential since the molecule lifetime in the lattice without atom removal is only a few ms, which we attribute to tunneling of the K atoms that enables molecule-atom chemical reactions~\cite{chemistry}.  After holding the ground-state molecules for 40 ms in the lattice, we take $\textit{in situ}$ images of the molecule distribution by reversing the STIRAP process, dissociating the Feshbach molecules, and then imaging the K atoms.  Figure \ref{fig:figfinal}C shows images of the ground-state molecules in the lattice for cases of both high and low conversion (the arrows indicate the regimes for the two images). The top image corresponds to starting with 2500 Rb atoms while the bottom image corresponds to starting with roughly 25000 Rb atoms.  The bottom image exhibits a central hole in the molecule distribution, which is consistent with the fact that the central lattice sites contained more than one Rb atom and therefore did not produce molecules.

For the higher conversion case, we perform a TF fit to the ground-state molecular distribution.   From the fit we find $7.9(5) \times 10^2$ molecules with a TF radius of $12.0(2)$ $\mu$m.  This gives $f_{\text{TF}}=0.27(2)$.  As an alternative approach, we can determine the filling by comparing the width of the molecular cloud with that of our simulated $T=0$ Rb distribution and assuming a uniform conversion efficiency of Rb into molecules.  The molecules are best described by a distribution that corresponds to an initial Rb number of $3.2(4) \times 10^3$.  Taking the ratio of the measured number of molecules to this Rb number, we find $f_{\text{mol}}=0.25(4)$, which is consistent with $f_{\text{TF}}$.  
 From the product of the previous measurements, namely the Rb filling, $N_{\text{KRb*}}/N_{\text{Rb}}$, and the STIRAP efficiency, one might expect a ground-state molecule filling of $\sim 35\%$.  We attribute the lower measured filling to molecular loss caused by the atom removals.
  
Given the ac polarizability~\cite{neyenhuis} and mass of the ground-state molecules, a lattice with a depth of 25$E_{\text{R}}^{\text{Rb}}$corresponds to $62E_{\text{R}}^{\text{KRb}}$, where $E_{\text{R}}^{\text{KRb}}$ is the recoil energy for a KRb molecule. The tunneling rate for molecules is therefore negligible. In this case, the entropy per molecule can be estimated from the filling in the lattice, with some assumption about the shape of the distribution.  Our approach of creating molecules from overlapping Rb and K insulators likely leads to a molecular distribution that is much more homogenous than the alternative approach of adiabatically loading a Fermi gas of molecules into the lattice.  The K Fermi gas is homogeneous within the confines of the initial Rb single-shell MI, which should result in a relatively uniform molecular distribution. For an average filling of $f_{\text{mol}}$ in a uniform lattice, the entropy per particle is $-\frac{k_B}{f_{\text{mol}}} [f_{\text{mol}} \ln(f_{\text{mol}})+(1-f_{\text{mol}})\ln(1-f_{\text{mol}})]$, which is $2.2~k_B$ for $f_{\text{mol}}=25\%$ \cite{fillinguniform}.   
For comparison, to reach this entropy by adiabatically loading a Fermi gas into a lattice would require starting with a quantum degenerate gas of molecules at $T/T_F=0.25$.

We find that the employment of dual atomic insulators has produced ground-state molecules in a 3D lattice with a very low entropy and a filling that is near the percolation threshold. Under this condition, the system of polar molecules in a 3D lattice is well connected and well suited for experiments probing the propagation of spin excitations in a system with long-range dipolar interactions. While the overall number of molecules now is lower than in previous work \cite{bonature}, the system realized here is appropriately sized for imaging with recently developed quantum gas microscope techniques \cite{greinermicroscope, blochmicroscope, chinmott}. More generally, this work elucidates the many challenges in, and extends the experimental toolbox for, synthesizing ultracold molecule systems that can realize novel quantum many-body behavior.

\begin{acknowledgments}
We thank Zhengkun Fu for experimental assistance, and Michael Wall, Arghavan Safavi, Kaden Hazzard, and Ana Maria Rey for many useful discussions.  We acknowledge funding from NIST, AFOSR-MURI, ARO-MURI, and NSF grant number 1125844.  J. P. C. is funded with an NDSEG graduate fellowship.
\end{acknowledgments}

\bibliography{finalbib}

\begin{thebibliography}{46}%
\makeatletter
\providecommand \@ifxundefined [1]{%
 \@ifx{#1\undefined}
}%
\providecommand \@ifnum [1]{%
 \ifnum #1\expandafter \@firstoftwo
 \else \expandafter \@secondoftwo
 \fi
}%
\providecommand \@ifx [1]{%
 \ifx #1\expandafter \@firstoftwo
 \else \expandafter \@secondoftwo
 \fi
}%
\providecommand \natexlab [1]{#1}%
\providecommand \enquote  [1]{``#1''}%
\providecommand \bibnamefont  [1]{#1}%
\providecommand \bibfnamefont [1]{#1}%
\providecommand \citenamefont [1]{#1}%
\providecommand \href@noop [0]{\@secondoftwo}%
\providecommand \href [0]{\begingroup \@sanitize@url \@href}%
\providecommand \@href[1]{\@@startlink{#1}\@@href}%
\providecommand \@@href[1]{\endgroup#1\@@endlink}%
\providecommand \@sanitize@url [0]{\catcode `\\12\catcode `\$12\catcode
  `\&12\catcode `\#12\catcode `\^12\catcode `\_12\catcode `\%12\relax}%
\providecommand \@@startlink[1]{}%
\providecommand \@@endlink[0]{}%
\providecommand \url  [0]{\begingroup\@sanitize@url \@url }%
\providecommand \@url [1]{\endgroup\@href {#1}{\urlprefix }}%
\providecommand \urlprefix  [0]{URL }%
\providecommand \Eprint [0]{\href }%
\providecommand \doibase [0]{http://dx.doi.org/}%
\providecommand \selectlanguage [0]{\@gobble}%
\providecommand \bibinfo  [0]{\@secondoftwo}%
\providecommand \bibfield  [0]{\@secondoftwo}%
\providecommand \translation [1]{[#1]}%
\providecommand \BibitemOpen [0]{}%
\providecommand \bibitemStop [0]{}%
\providecommand \bibitemNoStop [0]{.\EOS\space}%
\providecommand \EOS [0]{\spacefactor3000\relax}%
\providecommand \BibitemShut  [1]{\csname bibitem#1\endcsname}%
\let\auto@bib@innerbib\@empty
\bibitem [{\citenamefont {Micheli}\ \emph {et~al.}(2006)\citenamefont
  {Micheli}, \citenamefont {Brennen},\ and\ \citenamefont {Zoller}}]{toolbox}%
  \BibitemOpen
  \bibfield  {author} {\bibinfo {author} {\bibfnamefont {A.}~\bibnamefont
  {Micheli}}, \bibinfo {author} {\bibfnamefont {G.~K.}\ \bibnamefont
  {Brennen}}, \ and\ \bibinfo {author} {\bibfnamefont {P.}~\bibnamefont
  {Zoller}},\ }\href {http://dx.doi.org/10.1038/nphys287} {\bibfield  {journal}
  {\bibinfo  {journal} {Nat. Phys.}\ }\textbf {\bibinfo {volume} {2}},\
  \bibinfo {pages} {341} (\bibinfo {year} {2006})}\BibitemShut {NoStop}%
\bibitem [{\citenamefont {Barnett}\ \emph {et~al.}(2006)\citenamefont
  {Barnett}, \citenamefont {Petrov}, \citenamefont {Lukin},\ and\ \citenamefont
  {Demler}}]{demlermagnetism}%
  \BibitemOpen
  \bibfield  {author} {\bibinfo {author} {\bibfnamefont {R.}~\bibnamefont
  {Barnett}}, \bibinfo {author} {\bibfnamefont {D.}~\bibnamefont {Petrov}},
  \bibinfo {author} {\bibfnamefont {M.}~\bibnamefont {Lukin}}, \ and\ \bibinfo
  {author} {\bibfnamefont {E.}~\bibnamefont {Demler}},\ }\href {\doibase
  10.1103/PhysRevLett.96.190401} {\bibfield  {journal} {\bibinfo  {journal}
  {Phys. Rev. Lett.}\ }\textbf {\bibinfo {volume} {96}},\ \bibinfo {pages}
  {190401} (\bibinfo {year} {2006})}\BibitemShut {NoStop}%
\bibitem [{\citenamefont {Wall}\ \emph {et~al.}(2014)\citenamefont {Wall},
  \citenamefont {Hazzard},\ and\ \citenamefont {Rey}}]{qmagreview}%
  \BibitemOpen
  \bibfield  {author} {\bibinfo {author} {\bibfnamefont {M.~L.}\ \bibnamefont
  {Wall}}, \bibinfo {author} {\bibfnamefont {K.~R.~A.}\ \bibnamefont
  {Hazzard}}, \ and\ \bibinfo {author} {\bibfnamefont {A.~M.}\ \bibnamefont
  {Rey}},\ }\href@noop {} {\bibfield  {journal} {\bibinfo  {journal}
  {arXiv1406.4758v1}\ } (\bibinfo {year} {2014})}\BibitemShut {NoStop}%
\bibitem [{\citenamefont {Takekoshi}\ \emph {et~al.}(2014)\citenamefont
  {Takekoshi}, \citenamefont {Reichs\"ollner}, \citenamefont {Schindewolf},
  \citenamefont {Hutson}, \citenamefont {Le~Sueur}, \citenamefont {Dulieu},
  \citenamefont {Ferlaino}, \citenamefont {Grimm},\ and\ \citenamefont
  {N\"agerl}}]{rbcs}%
  \BibitemOpen
  \bibfield  {author} {\bibinfo {author} {\bibfnamefont {T.}~\bibnamefont
  {Takekoshi}}, \bibinfo {author} {\bibfnamefont {L.}~\bibnamefont
  {Reichs\"ollner}}, \bibinfo {author} {\bibfnamefont {A.}~\bibnamefont
  {Schindewolf}}, \bibinfo {author} {\bibfnamefont {J.~M.}\ \bibnamefont
  {Hutson}}, \bibinfo {author} {\bibfnamefont {C.~R.}\ \bibnamefont
  {Le~Sueur}}, \bibinfo {author} {\bibfnamefont {O.}~\bibnamefont {Dulieu}},
  \bibinfo {author} {\bibfnamefont {F.}~\bibnamefont {Ferlaino}}, \bibinfo
  {author} {\bibfnamefont {R.}~\bibnamefont {Grimm}}, \ and\ \bibinfo {author}
  {\bibfnamefont {H.-C.}\ \bibnamefont {N\"agerl}},\ }\href {\doibase
  10.1103/PhysRevLett.113.205301} {\bibfield  {journal} {\bibinfo  {journal}
  {Phys. Rev. Lett.}\ }\textbf {\bibinfo {volume} {113}},\ \bibinfo {pages}
  {205301} (\bibinfo {year} {2014})}\BibitemShut {NoStop}%
\bibitem [{\citenamefont {Ni}\ \emph {et~al.}(2008)\citenamefont {Ni},
  \citenamefont {Ospelkaus}, \citenamefont {de~Miranda}, \citenamefont {Pe'er},
  \citenamefont {Neyenhuis}, \citenamefont {Zirbel}, \citenamefont
  {Kotochigova}, \citenamefont {Julienne}, \citenamefont {Jin},\ and\
  \citenamefont {Ye}}]{kangkuen}%
  \BibitemOpen
  \bibfield  {author} {\bibinfo {author} {\bibfnamefont {K.-K.}\ \bibnamefont
  {Ni}}, \bibinfo {author} {\bibfnamefont {S.}~\bibnamefont {Ospelkaus}},
  \bibinfo {author} {\bibfnamefont {M.~H.~G.}\ \bibnamefont {de~Miranda}},
  \bibinfo {author} {\bibfnamefont {A.}~\bibnamefont {Pe'er}}, \bibinfo
  {author} {\bibfnamefont {B.}~\bibnamefont {Neyenhuis}}, \bibinfo {author}
  {\bibfnamefont {J.~J.}\ \bibnamefont {Zirbel}}, \bibinfo {author}
  {\bibfnamefont {S.}~\bibnamefont {Kotochigova}}, \bibinfo {author}
  {\bibfnamefont {P.~S.}\ \bibnamefont {Julienne}}, \bibinfo {author}
  {\bibfnamefont {D.~S.}\ \bibnamefont {Jin}}, \ and\ \bibinfo {author}
  {\bibfnamefont {J.}~\bibnamefont {Ye}},\ }\href {\doibase
  10.1126/science.1163861} {\bibfield  {journal} {\bibinfo  {journal}
  {Science}\ }\textbf {\bibinfo {volume} {322}},\ \bibinfo {pages} {231}
  (\bibinfo {year} {2008})}\BibitemShut {NoStop}%
\bibitem [{\citenamefont {Ospelkaus}\ \emph {et~al.}(2010)\citenamefont
  {Ospelkaus}, \citenamefont {Ni}, \citenamefont {Wang}, \citenamefont
  {de~Miranda}, \citenamefont {Neyenhuis}, \citenamefont {Qu\'{e}m\'{e}ner},
  \citenamefont {Julienne}, \citenamefont {Bohn}, \citenamefont {Jin},\ and\
  \citenamefont {Ye}}]{chemistry}%
  \BibitemOpen
  \bibfield  {author} {\bibinfo {author} {\bibfnamefont {S.}~\bibnamefont
  {Ospelkaus}}, \bibinfo {author} {\bibfnamefont {K.-K.}\ \bibnamefont {Ni}},
  \bibinfo {author} {\bibfnamefont {D.}~\bibnamefont {Wang}}, \bibinfo {author}
  {\bibfnamefont {M.~H.~G.}\ \bibnamefont {de~Miranda}}, \bibinfo {author}
  {\bibfnamefont {B.}~\bibnamefont {Neyenhuis}}, \bibinfo {author}
  {\bibfnamefont {G.}~\bibnamefont {Qu\'{e}m\'{e}ner}}, \bibinfo {author}
  {\bibfnamefont {P.~S.}\ \bibnamefont {Julienne}}, \bibinfo {author}
  {\bibfnamefont {J.~L.}\ \bibnamefont {Bohn}}, \bibinfo {author}
  {\bibfnamefont {D.~S.}\ \bibnamefont {Jin}}, \ and\ \bibinfo {author}
  {\bibfnamefont {J.}~\bibnamefont {Ye}},\ }\href {\doibase
  10.1126/science.1184121} {\bibfield  {journal} {\bibinfo  {journal}
  {Science}\ }\textbf {\bibinfo {volume} {327}},\ \bibinfo {pages} {853}
  (\bibinfo {year} {2010})}\BibitemShut {NoStop}%
\bibitem [{\citenamefont {B\"uchler}\ \emph {et~al.}(2007)\citenamefont
  {B\"uchler}, \citenamefont {Demler}, \citenamefont {Lukin}, \citenamefont
  {Micheli}, \citenamefont {Prokof'ev}, \citenamefont {Pupillo},\ and\
  \citenamefont {Zoller}}]{buchlerpotential}%
  \BibitemOpen
  \bibfield  {author} {\bibinfo {author} {\bibfnamefont {H.~P.}\ \bibnamefont
  {B\"uchler}}, \bibinfo {author} {\bibfnamefont {E.}~\bibnamefont {Demler}},
  \bibinfo {author} {\bibfnamefont {M.}~\bibnamefont {Lukin}}, \bibinfo
  {author} {\bibfnamefont {A.}~\bibnamefont {Micheli}}, \bibinfo {author}
  {\bibfnamefont {N.}~\bibnamefont {Prokof'ev}}, \bibinfo {author}
  {\bibfnamefont {G.}~\bibnamefont {Pupillo}}, \ and\ \bibinfo {author}
  {\bibfnamefont {P.}~\bibnamefont {Zoller}},\ }\href {\doibase
  10.1103/PhysRevLett.98.060404} {\bibfield  {journal} {\bibinfo  {journal}
  {Phys. Rev. Lett.}\ }\textbf {\bibinfo {volume} {98}},\ \bibinfo {pages}
  {060404} (\bibinfo {year} {2007})}\BibitemShut {NoStop}%
\bibitem [{\citenamefont {de~Miranda}\ \emph {et~al.}(2011)\citenamefont
  {de~Miranda}, \citenamefont {Chotia}, \citenamefont {Neyenhuis},
  \citenamefont {Wang}, \citenamefont {Qu\'{e}m\'{e}ner}, \citenamefont
  {Ospelkaus}, \citenamefont {Bohn}, \citenamefont {Ye},\ and\ \citenamefont
  {Jin}}]{marcio}%
  \BibitemOpen
  \bibfield  {author} {\bibinfo {author} {\bibfnamefont {M.~H.~G.}\
  \bibnamefont {de~Miranda}}, \bibinfo {author} {\bibfnamefont
  {A.}~\bibnamefont {Chotia}}, \bibinfo {author} {\bibfnamefont
  {B.}~\bibnamefont {Neyenhuis}}, \bibinfo {author} {\bibfnamefont
  {D.}~\bibnamefont {Wang}}, \bibinfo {author} {\bibfnamefont {G.}~\bibnamefont
  {Qu\'{e}m\'{e}ner}}, \bibinfo {author} {\bibfnamefont {S.}~\bibnamefont
  {Ospelkaus}}, \bibinfo {author} {\bibfnamefont {J.~L.}\ \bibnamefont {Bohn}},
  \bibinfo {author} {\bibfnamefont {J.}~\bibnamefont {Ye}}, \ and\ \bibinfo
  {author} {\bibfnamefont {D.~S.}\ \bibnamefont {Jin}},\ }\href
  {http://dx.doi.org/10.1038/nphys1939} {\bibfield  {journal} {\bibinfo
  {journal} {Nat. Phys.}\ }\textbf {\bibinfo {volume} {7}},\ \bibinfo {pages}
  {502} (\bibinfo {year} {2011})}\BibitemShut {NoStop}%
\bibitem [{\citenamefont {Chotia}\ \emph {et~al.}(2012)\citenamefont {Chotia},
  \citenamefont {Neyenhuis}, \citenamefont {Moses}, \citenamefont {Yan},
  \citenamefont {Covey}, \citenamefont {Foss-Feig}, \citenamefont {Rey},
  \citenamefont {Jin},\ and\ \citenamefont {Ye}}]{amodsenprl}%
  \BibitemOpen
  \bibfield  {author} {\bibinfo {author} {\bibfnamefont {A.}~\bibnamefont
  {Chotia}}, \bibinfo {author} {\bibfnamefont {B.}~\bibnamefont {Neyenhuis}},
  \bibinfo {author} {\bibfnamefont {S.~A.}\ \bibnamefont {Moses}}, \bibinfo
  {author} {\bibfnamefont {B.}~\bibnamefont {Yan}}, \bibinfo {author}
  {\bibfnamefont {J.~P.}\ \bibnamefont {Covey}}, \bibinfo {author}
  {\bibfnamefont {M.}~\bibnamefont {Foss-Feig}}, \bibinfo {author}
  {\bibfnamefont {A.~M.}\ \bibnamefont {Rey}}, \bibinfo {author} {\bibfnamefont
  {D.~S.}\ \bibnamefont {Jin}}, \ and\ \bibinfo {author} {\bibfnamefont
  {J.}~\bibnamefont {Ye}},\ }\href {\doibase 10.1103/PhysRevLett.108.080405}
  {\bibfield  {journal} {\bibinfo  {journal} {Phys. Rev. Lett.}\ }\textbf
  {\bibinfo {volume} {108}},\ \bibinfo {pages} {080405} (\bibinfo {year}
  {2012})}\BibitemShut {NoStop}%
\bibitem [{\citenamefont {Bloch}\ \emph {et~al.}(2012)\citenamefont {Bloch},
  \citenamefont {Dalibard},\ and\ \citenamefont
  {Nascimbene}}]{qmagatomsreview}%
  \BibitemOpen
  \bibfield  {author} {\bibinfo {author} {\bibfnamefont {I.}~\bibnamefont
  {Bloch}}, \bibinfo {author} {\bibfnamefont {J.}~\bibnamefont {Dalibard}}, \
  and\ \bibinfo {author} {\bibfnamefont {S.}~\bibnamefont {Nascimbene}},\
  }\href {http://dx.doi.org/10.1038/nphys2259} {\bibfield  {journal} {\bibinfo
  {journal} {Nat. Phys.}\ }\textbf {\bibinfo {volume} {8}},\ \bibinfo {pages}
  {267} (\bibinfo {year} {2012})}\BibitemShut {NoStop}%
\bibitem [{\citenamefont {Hazzard}\ \emph {et~al.}(2013)\citenamefont
  {Hazzard}, \citenamefont {Manmana}, \citenamefont {Foss-Feig},\ and\
  \citenamefont {Rey}}]{kadenproposal}%
  \BibitemOpen
  \bibfield  {author} {\bibinfo {author} {\bibfnamefont {K.~R.~A.}\
  \bibnamefont {Hazzard}}, \bibinfo {author} {\bibfnamefont {S.~R.}\
  \bibnamefont {Manmana}}, \bibinfo {author} {\bibfnamefont {M.}~\bibnamefont
  {Foss-Feig}}, \ and\ \bibinfo {author} {\bibfnamefont {A.~M.}\ \bibnamefont
  {Rey}},\ }\href {\doibase 10.1103/PhysRevLett.110.075301} {\bibfield
  {journal} {\bibinfo  {journal} {Phys. Rev. Lett.}\ }\textbf {\bibinfo
  {volume} {110}},\ \bibinfo {pages} {075301} (\bibinfo {year}
  {2013})}\BibitemShut {NoStop}%
\bibitem [{\citenamefont {Yan}\ \emph {et~al.}(2013)\citenamefont {Yan},
  \citenamefont {Moses}, \citenamefont {Gadway}, \citenamefont {Covey},
  \citenamefont {Hazzard}, \citenamefont {Rey}, \citenamefont {Jin},\ and\
  \citenamefont {Ye}}]{bonature}%
  \BibitemOpen
  \bibfield  {author} {\bibinfo {author} {\bibfnamefont {B.}~\bibnamefont
  {Yan}}, \bibinfo {author} {\bibfnamefont {S.~A.}\ \bibnamefont {Moses}},
  \bibinfo {author} {\bibfnamefont {B.}~\bibnamefont {Gadway}}, \bibinfo
  {author} {\bibfnamefont {J.~P.}\ \bibnamefont {Covey}}, \bibinfo {author}
  {\bibfnamefont {K.~R.~A.}\ \bibnamefont {Hazzard}}, \bibinfo {author}
  {\bibfnamefont {A.~M.}\ \bibnamefont {Rey}}, \bibinfo {author} {\bibfnamefont
  {D.~S.}\ \bibnamefont {Jin}}, \ and\ \bibinfo {author} {\bibfnamefont
  {J.}~\bibnamefont {Ye}},\ }\href {http://dx.doi.org/10.1038/nature12483}
  {\bibfield  {journal} {\bibinfo  {journal} {Nature}\ }\textbf {\bibinfo
  {volume} {501}},\ \bibinfo {pages} {521} (\bibinfo {year}
  {2013})}\BibitemShut {NoStop}%
\bibitem [{\citenamefont {Hazzard}\ \emph
  {et~al.}(2014{\natexlab{a}})\citenamefont {Hazzard}, \citenamefont {Gadway},
  \citenamefont {Foss-Feig}, \citenamefont {Yan}, \citenamefont {Moses},
  \citenamefont {Covey}, \citenamefont {Yao}, \citenamefont {Lukin},
  \citenamefont {Ye}, \citenamefont {Jin},\ and\ \citenamefont
  {Rey}}]{mbdynamics}%
  \BibitemOpen
  \bibfield  {author} {\bibinfo {author} {\bibfnamefont {K.~R.~A.}\
  \bibnamefont {Hazzard}}, \bibinfo {author} {\bibfnamefont {B.}~\bibnamefont
  {Gadway}}, \bibinfo {author} {\bibfnamefont {M.}~\bibnamefont {Foss-Feig}},
  \bibinfo {author} {\bibfnamefont {B.}~\bibnamefont {Yan}}, \bibinfo {author}
  {\bibfnamefont {S.~A.}\ \bibnamefont {Moses}}, \bibinfo {author}
  {\bibfnamefont {J.~P.}\ \bibnamefont {Covey}}, \bibinfo {author}
  {\bibfnamefont {N.~Y.}\ \bibnamefont {Yao}}, \bibinfo {author} {\bibfnamefont
  {M.~D.}\ \bibnamefont {Lukin}}, \bibinfo {author} {\bibfnamefont
  {J.}~\bibnamefont {Ye}}, \bibinfo {author} {\bibfnamefont {D.~S.}\
  \bibnamefont {Jin}}, \ and\ \bibinfo {author} {\bibfnamefont {A.~M.}\
  \bibnamefont {Rey}},\ }\href {\doibase 10.1103/PhysRevLett.113.195302}
  {\bibfield  {journal} {\bibinfo  {journal} {Phys. Rev. Lett.}\ }\textbf
  {\bibinfo {volume} {113}},\ \bibinfo {pages} {195302} (\bibinfo {year}
  {2014}{\natexlab{a}})}\BibitemShut {NoStop}%
\bibitem [{\citenamefont {Gorshkov}\ \emph {et~al.}(2011)\citenamefont
  {Gorshkov}, \citenamefont {Manmana}, \citenamefont {Chen}, \citenamefont
  {Ye}, \citenamefont {Demler}, \citenamefont {Lukin},\ and\ \citenamefont
  {Rey}}]{tjvw}%
  \BibitemOpen
  \bibfield  {author} {\bibinfo {author} {\bibfnamefont {A.~V.}\ \bibnamefont
  {Gorshkov}}, \bibinfo {author} {\bibfnamefont {S.~R.}\ \bibnamefont
  {Manmana}}, \bibinfo {author} {\bibfnamefont {G.}~\bibnamefont {Chen}},
  \bibinfo {author} {\bibfnamefont {J.}~\bibnamefont {Ye}}, \bibinfo {author}
  {\bibfnamefont {E.}~\bibnamefont {Demler}}, \bibinfo {author} {\bibfnamefont
  {M.~D.}\ \bibnamefont {Lukin}}, \ and\ \bibinfo {author} {\bibfnamefont
  {A.~M.}\ \bibnamefont {Rey}},\ }\href {\doibase
  10.1103/PhysRevLett.107.115301} {\bibfield  {journal} {\bibinfo  {journal}
  {Phys. Rev. Lett.}\ }\textbf {\bibinfo {volume} {107}},\ \bibinfo {pages}
  {115301} (\bibinfo {year} {2011})}\BibitemShut {NoStop}%
\bibitem [{\citenamefont {de~Paz}\ \emph {et~al.}(2013)\citenamefont {de~Paz},
  \citenamefont {Sharma}, \citenamefont {Chotia}, \citenamefont {Mar\'echal},
  \citenamefont {Huckans}, \citenamefont {Pedri}, \citenamefont {Santos},
  \citenamefont {Gorceix}, \citenamefont {Vernac},\ and\ \citenamefont
  {Laburthe-Tolra}}]{crmagnetism}%
  \BibitemOpen
  \bibfield  {author} {\bibinfo {author} {\bibfnamefont {A.}~\bibnamefont
  {de~Paz}}, \bibinfo {author} {\bibfnamefont {A.}~\bibnamefont {Sharma}},
  \bibinfo {author} {\bibfnamefont {A.}~\bibnamefont {Chotia}}, \bibinfo
  {author} {\bibfnamefont {E.}~\bibnamefont {Mar\'echal}}, \bibinfo {author}
  {\bibfnamefont {J.~H.}\ \bibnamefont {Huckans}}, \bibinfo {author}
  {\bibfnamefont {P.}~\bibnamefont {Pedri}}, \bibinfo {author} {\bibfnamefont
  {L.}~\bibnamefont {Santos}}, \bibinfo {author} {\bibfnamefont
  {O.}~\bibnamefont {Gorceix}}, \bibinfo {author} {\bibfnamefont
  {L.}~\bibnamefont {Vernac}}, \ and\ \bibinfo {author} {\bibfnamefont
  {B.}~\bibnamefont {Laburthe-Tolra}},\ }\href {\doibase
  10.1103/PhysRevLett.111.185305} {\bibfield  {journal} {\bibinfo  {journal}
  {Phys. Rev. Lett.}\ }\textbf {\bibinfo {volume} {111}},\ \bibinfo {pages}
  {185305} (\bibinfo {year} {2013})}\BibitemShut {NoStop}%
\bibitem [{\citenamefont {Fukuhara}\ \emph {et~al.}(2013)\citenamefont
  {Fukuhara}, \citenamefont {Kantian}, \citenamefont {Endres}, \citenamefont
  {Cheneau}, \citenamefont {Schausz}, \citenamefont {Hild}, \citenamefont
  {Bellem}, \citenamefont {Schollwock}, \citenamefont {Giamarchi},
  \citenamefont {Gross}, \citenamefont {Bloch},\ and\ \citenamefont
  {Kuhr}}]{munichspin}%
  \BibitemOpen
  \bibfield  {author} {\bibinfo {author} {\bibfnamefont {T.}~\bibnamefont
  {Fukuhara}}, \bibinfo {author} {\bibfnamefont {A.}~\bibnamefont {Kantian}},
  \bibinfo {author} {\bibfnamefont {M.}~\bibnamefont {Endres}}, \bibinfo
  {author} {\bibfnamefont {M.}~\bibnamefont {Cheneau}}, \bibinfo {author}
  {\bibfnamefont {P.}~\bibnamefont {Schausz}}, \bibinfo {author} {\bibfnamefont
  {S.}~\bibnamefont {Hild}}, \bibinfo {author} {\bibfnamefont {D.}~\bibnamefont
  {Bellem}}, \bibinfo {author} {\bibfnamefont {U.}~\bibnamefont {Schollwock}},
  \bibinfo {author} {\bibfnamefont {T.}~\bibnamefont {Giamarchi}}, \bibinfo
  {author} {\bibfnamefont {C.}~\bibnamefont {Gross}}, \bibinfo {author}
  {\bibfnamefont {I.}~\bibnamefont {Bloch}}, \ and\ \bibinfo {author}
  {\bibfnamefont {S.}~\bibnamefont {Kuhr}},\ }\href
  {http://dx.doi.org/10.1038/nphys2561} {\bibfield  {journal} {\bibinfo
  {journal} {Nat. Phys.}\ }\textbf {\bibinfo {volume} {9}},\ \bibinfo {pages}
  {235} (\bibinfo {year} {2013})}\BibitemShut {NoStop}%
\bibitem [{\citenamefont {Richerme}\ \emph {et~al.}(2014)\citenamefont
  {Richerme}, \citenamefont {Gong}, \citenamefont {Lee}, \citenamefont {Senko},
  \citenamefont {Smith}, \citenamefont {Foss-Feig}, \citenamefont {Michalakis},
  \citenamefont {Gorshkov},\ and\ \citenamefont {Monroe}}]{monroe}%
  \BibitemOpen
  \bibfield  {author} {\bibinfo {author} {\bibfnamefont {P.}~\bibnamefont
  {Richerme}}, \bibinfo {author} {\bibfnamefont {Z.-X.}\ \bibnamefont {Gong}},
  \bibinfo {author} {\bibfnamefont {A.}~\bibnamefont {Lee}}, \bibinfo {author}
  {\bibfnamefont {C.}~\bibnamefont {Senko}}, \bibinfo {author} {\bibfnamefont
  {J.}~\bibnamefont {Smith}}, \bibinfo {author} {\bibfnamefont
  {M.}~\bibnamefont {Foss-Feig}}, \bibinfo {author} {\bibfnamefont
  {S.}~\bibnamefont {Michalakis}}, \bibinfo {author} {\bibfnamefont {A.~V.}\
  \bibnamefont {Gorshkov}}, \ and\ \bibinfo {author} {\bibfnamefont
  {C.}~\bibnamefont {Monroe}},\ }\href {http://dx.doi.org/10.1038/nature13450}
  {\bibfield  {journal} {\bibinfo  {journal} {Nature}\ }\textbf {\bibinfo
  {volume} {511}},\ \bibinfo {pages} {198} (\bibinfo {year}
  {2014})}\BibitemShut {NoStop}%
\bibitem [{\citenamefont {Jurcevic}\ \emph {et~al.}(2014)\citenamefont
  {Jurcevic}, \citenamefont {Lanyon}, \citenamefont {Hauke}, \citenamefont
  {Hempel}, \citenamefont {Zoller}, \citenamefont {Blatt},\ and\ \citenamefont
  {Roos}}]{blatt}%
  \BibitemOpen
  \bibfield  {author} {\bibinfo {author} {\bibfnamefont {P.}~\bibnamefont
  {Jurcevic}}, \bibinfo {author} {\bibfnamefont {B.~P.}\ \bibnamefont
  {Lanyon}}, \bibinfo {author} {\bibfnamefont {P.}~\bibnamefont {Hauke}},
  \bibinfo {author} {\bibfnamefont {C.}~\bibnamefont {Hempel}}, \bibinfo
  {author} {\bibfnamefont {P.}~\bibnamefont {Zoller}}, \bibinfo {author}
  {\bibfnamefont {R.}~\bibnamefont {Blatt}}, \ and\ \bibinfo {author}
  {\bibfnamefont {C.~F.}\ \bibnamefont {Roos}},\ }\href
  {http://dx.doi.org/10.1038/nature13461} {\bibfield  {journal} {\bibinfo
  {journal} {Nature}\ }\textbf {\bibinfo {volume} {511}},\ \bibinfo {pages}
  {202} (\bibinfo {year} {2014})}\BibitemShut {NoStop}%
\bibitem [{\citenamefont {Cai}\ \emph {et~al.}(2013)\citenamefont {Cai},
  \citenamefont {Retzker}, \citenamefont {Jelezko},\ and\ \citenamefont
  {Plenio}}]{diamond}%
  \BibitemOpen
  \bibfield  {author} {\bibinfo {author} {\bibfnamefont {J.}~\bibnamefont
  {Cai}}, \bibinfo {author} {\bibfnamefont {A.}~\bibnamefont {Retzker}},
  \bibinfo {author} {\bibfnamefont {F.}~\bibnamefont {Jelezko}}, \ and\
  \bibinfo {author} {\bibfnamefont {M.~B.}\ \bibnamefont {Plenio}},\ }\href
  {http://dx.doi.org/10.1038/nphys2519} {\bibfield  {journal} {\bibinfo
  {journal} {Nat. Phys.}\ }\textbf {\bibinfo {volume} {9}},\ \bibinfo {pages}
  {168} (\bibinfo {year} {2013})}\BibitemShut {NoStop}%
\bibitem [{\citenamefont {Ravets}\ \emph {et~al.}(2014)\citenamefont {Ravets},
  \citenamefont {Labuhn}, \citenamefont {Barredo}, \citenamefont {Beguin},
  \citenamefont {Lahaye},\ and\ \citenamefont {Browaeys}}]{rydberg}%
  \BibitemOpen
  \bibfield  {author} {\bibinfo {author} {\bibfnamefont {S.}~\bibnamefont
  {Ravets}}, \bibinfo {author} {\bibfnamefont {H.}~\bibnamefont {Labuhn}},
  \bibinfo {author} {\bibfnamefont {D.}~\bibnamefont {Barredo}}, \bibinfo
  {author} {\bibfnamefont {L.}~\bibnamefont {Beguin}}, \bibinfo {author}
  {\bibfnamefont {T.}~\bibnamefont {Lahaye}}, \ and\ \bibinfo {author}
  {\bibfnamefont {A.}~\bibnamefont {Browaeys}},\ }\href
  {http://dx.doi.org/10.1038/nphys3119} {\bibfield  {journal} {\bibinfo
  {journal} {Nat. Phys.}\ }\textbf {\bibinfo {volume} {10}},\ \bibinfo {pages}
  {914} (\bibinfo {year} {2014})}\BibitemShut {NoStop}%
\bibitem [{\citenamefont {Hazzard}\ \emph
  {et~al.}(2014{\natexlab{b}})\citenamefont {Hazzard}, \citenamefont {van~den
  Worm}, \citenamefont {Foss-Feig}, \citenamefont {Manmana}, \citenamefont
  {Dalla~Torre}, \citenamefont {Pfau}, \citenamefont {Kastner},\ and\
  \citenamefont {Rey}}]{correlations}%
  \BibitemOpen
  \bibfield  {author} {\bibinfo {author} {\bibfnamefont {K.~R.~A.}\
  \bibnamefont {Hazzard}}, \bibinfo {author} {\bibfnamefont {M.}~\bibnamefont
  {van~den Worm}}, \bibinfo {author} {\bibfnamefont {M.}~\bibnamefont
  {Foss-Feig}}, \bibinfo {author} {\bibfnamefont {S.~R.}\ \bibnamefont
  {Manmana}}, \bibinfo {author} {\bibfnamefont {E.~G.}\ \bibnamefont
  {Dalla~Torre}}, \bibinfo {author} {\bibfnamefont {T.}~\bibnamefont {Pfau}},
  \bibinfo {author} {\bibfnamefont {M.}~\bibnamefont {Kastner}}, \ and\
  \bibinfo {author} {\bibfnamefont {A.~M.}\ \bibnamefont {Rey}},\ }\href
  {\doibase 10.1103/PhysRevA.90.063622} {\bibfield  {journal} {\bibinfo
  {journal} {Phys. Rev. A}\ }\textbf {\bibinfo {volume} {90}},\ \bibinfo
  {pages} {063622} (\bibinfo {year} {2014}{\natexlab{b}})}\BibitemShut
  {NoStop}%
\bibitem [{\citenamefont {Schachenmayer}\ \emph {et~al.}(2013)\citenamefont
  {Schachenmayer}, \citenamefont {Lanyon}, \citenamefont {Roos},\ and\
  \citenamefont {Daley}}]{johannesent}%
  \BibitemOpen
  \bibfield  {author} {\bibinfo {author} {\bibfnamefont {J.}~\bibnamefont
  {Schachenmayer}}, \bibinfo {author} {\bibfnamefont {B.~P.}\ \bibnamefont
  {Lanyon}}, \bibinfo {author} {\bibfnamefont {C.~F.}\ \bibnamefont {Roos}}, \
  and\ \bibinfo {author} {\bibfnamefont {A.~J.}\ \bibnamefont {Daley}},\ }\href
  {\doibase 10.1103/PhysRevX.3.031015} {\bibfield  {journal} {\bibinfo
  {journal} {Phys. Rev. X}\ }\textbf {\bibinfo {volume} {3}},\ \bibinfo {pages}
  {031015} (\bibinfo {year} {2013})}\BibitemShut {NoStop}%
\bibitem [{\citenamefont {Yao}\ \emph {et~al.}(2014)\citenamefont {Yao},
  \citenamefont {Laumann}, \citenamefont {Gopalakrishnan}, \citenamefont
  {Knap}, \citenamefont {M\"uller}, \citenamefont {Demler},\ and\ \citenamefont
  {Lukin}}]{mblocalization}%
  \BibitemOpen
  \bibfield  {author} {\bibinfo {author} {\bibfnamefont {N.~Y.}\ \bibnamefont
  {Yao}}, \bibinfo {author} {\bibfnamefont {C.~R.}\ \bibnamefont {Laumann}},
  \bibinfo {author} {\bibfnamefont {S.}~\bibnamefont {Gopalakrishnan}},
  \bibinfo {author} {\bibfnamefont {M.}~\bibnamefont {Knap}}, \bibinfo {author}
  {\bibfnamefont {M.}~\bibnamefont {M\"uller}}, \bibinfo {author}
  {\bibfnamefont {E.~A.}\ \bibnamefont {Demler}}, \ and\ \bibinfo {author}
  {\bibfnamefont {M.~D.}\ \bibnamefont {Lukin}},\ }\href {\doibase
  10.1103/PhysRevLett.113.243002} {\bibfield  {journal} {\bibinfo  {journal}
  {Phys. Rev. Lett.}\ }\textbf {\bibinfo {volume} {113}},\ \bibinfo {pages}
  {243002} (\bibinfo {year} {2014})}\BibitemShut {NoStop}%
\bibitem [{\citenamefont {Kuns}\ \emph {et~al.}(2011)\citenamefont {Kuns},
  \citenamefont {Rey},\ and\ \citenamefont {Gorshkov}}]{dwave}%
  \BibitemOpen
  \bibfield  {author} {\bibinfo {author} {\bibfnamefont {K.~A.}\ \bibnamefont
  {Kuns}}, \bibinfo {author} {\bibfnamefont {A.~M.}\ \bibnamefont {Rey}}, \
  and\ \bibinfo {author} {\bibfnamefont {A.~V.}\ \bibnamefont {Gorshkov}},\
  }\href {\doibase 10.1103/PhysRevA.84.063639} {\bibfield  {journal} {\bibinfo
  {journal} {Phys. Rev. A}\ }\textbf {\bibinfo {volume} {84}},\ \bibinfo
  {pages} {063639} (\bibinfo {year} {2011})}\BibitemShut {NoStop}%
\bibitem [{\citenamefont {Knap}\ \emph {et~al.}(2012)\citenamefont {Knap},
  \citenamefont {Berg}, \citenamefont {Ganahl},\ and\ \citenamefont
  {Demler}}]{wigner}%
  \BibitemOpen
  \bibfield  {author} {\bibinfo {author} {\bibfnamefont {M.}~\bibnamefont
  {Knap}}, \bibinfo {author} {\bibfnamefont {E.}~\bibnamefont {Berg}}, \bibinfo
  {author} {\bibfnamefont {M.}~\bibnamefont {Ganahl}}, \ and\ \bibinfo {author}
  {\bibfnamefont {E.}~\bibnamefont {Demler}},\ }\href {\doibase
  10.1103/PhysRevB.86.064501} {\bibfield  {journal} {\bibinfo  {journal} {Phys.
  Rev. B}\ }\textbf {\bibinfo {volume} {86}},\ \bibinfo {pages} {064501}
  (\bibinfo {year} {2012})}\BibitemShut {NoStop}%
\bibitem [{\citenamefont {Manmana}\ \emph {et~al.}(2013)\citenamefont
  {Manmana}, \citenamefont {Stoudenmire}, \citenamefont {Hazzard},
  \citenamefont {Rey},\ and\ \citenamefont {Gorshkov}}]{salvatoretopology}%
  \BibitemOpen
  \bibfield  {author} {\bibinfo {author} {\bibfnamefont {S.~R.}\ \bibnamefont
  {Manmana}}, \bibinfo {author} {\bibfnamefont {E.~M.}\ \bibnamefont
  {Stoudenmire}}, \bibinfo {author} {\bibfnamefont {K.~R.~A.}\ \bibnamefont
  {Hazzard}}, \bibinfo {author} {\bibfnamefont {A.~M.}\ \bibnamefont {Rey}}, \
  and\ \bibinfo {author} {\bibfnamefont {A.~V.}\ \bibnamefont {Gorshkov}},\
  }\href {\doibase 10.1103/PhysRevB.87.081106} {\bibfield  {journal} {\bibinfo
  {journal} {Phys. Rev. B}\ }\textbf {\bibinfo {volume} {87}},\ \bibinfo
  {pages} {081106} (\bibinfo {year} {2013})}\BibitemShut {NoStop}%
\bibitem [{\citenamefont {He}\ and\ \citenamefont
  {Hofstetter}(2011)}]{supersolid}%
  \BibitemOpen
  \bibfield  {author} {\bibinfo {author} {\bibfnamefont {L.}~\bibnamefont
  {He}}\ and\ \bibinfo {author} {\bibfnamefont {W.}~\bibnamefont
  {Hofstetter}},\ }\href {\doibase 10.1103/PhysRevA.83.053629} {\bibfield
  {journal} {\bibinfo  {journal} {Phys. Rev. A}\ }\textbf {\bibinfo {volume}
  {83}},\ \bibinfo {pages} {053629} (\bibinfo {year} {2011})}\BibitemShut
  {NoStop}%
\bibitem [{\citenamefont {Yao}\ \emph {et~al.}(2013)\citenamefont {Yao},
  \citenamefont {Gorshkov}, \citenamefont {Laumann}, \citenamefont {L\"auchli},
  \citenamefont {Ye},\ and\ \citenamefont {Lukin}}]{yaochern}%
  \BibitemOpen
  \bibfield  {author} {\bibinfo {author} {\bibfnamefont {N.~Y.}\ \bibnamefont
  {Yao}}, \bibinfo {author} {\bibfnamefont {A.~V.}\ \bibnamefont {Gorshkov}},
  \bibinfo {author} {\bibfnamefont {C.~R.}\ \bibnamefont {Laumann}}, \bibinfo
  {author} {\bibfnamefont {A.~M.}\ \bibnamefont {L\"auchli}}, \bibinfo {author}
  {\bibfnamefont {J.}~\bibnamefont {Ye}}, \ and\ \bibinfo {author}
  {\bibfnamefont {M.~D.}\ \bibnamefont {Lukin}},\ }\href {\doibase
  10.1103/PhysRevLett.110.185302} {\bibfield  {journal} {\bibinfo  {journal}
  {Phys. Rev. Lett.}\ }\textbf {\bibinfo {volume} {110}},\ \bibinfo {pages}
  {185302} (\bibinfo {year} {2013})}\BibitemShut {NoStop}%
\bibitem [{\citenamefont {Syzranov}\ \emph {et~al.}(2014)\citenamefont
  {Syzranov}, \citenamefont {Wall}, \citenamefont {Gurarie},\ and\
  \citenamefont {Rey}}]{socoupling}%
  \BibitemOpen
  \bibfield  {author} {\bibinfo {author} {\bibfnamefont {S.~V.}\ \bibnamefont
  {Syzranov}}, \bibinfo {author} {\bibfnamefont {M.~L.}\ \bibnamefont {Wall}},
  \bibinfo {author} {\bibfnamefont {V.}~\bibnamefont {Gurarie}}, \ and\
  \bibinfo {author} {\bibfnamefont {A.~M.}\ \bibnamefont {Rey}},\ }\href
  {http://dx.doi.org/10.1038/ncomms6391} {\bibfield  {journal} {\bibinfo
  {journal} {Nat. Commun.}\ }\textbf {\bibinfo {volume} {5}} (\bibinfo {year}
  {2014})}\BibitemShut {NoStop}%
\bibitem [{\citenamefont {Stauffer}\ and\ \citenamefont
  {Aharon}(1994)}]{percolation}%
  \BibitemOpen
  \bibfield  {author} {\bibinfo {author} {\bibfnamefont {D.}~\bibnamefont
  {Stauffer}}\ and\ \bibinfo {author} {\bibfnamefont {A.}~\bibnamefont
  {Aharon}},\ }\href@noop {} {\emph {\bibinfo {title} {Introduction to
  Percolation Theory: Revised Second Edition}}}\ (\bibinfo  {publisher} {Taylor
  and Francis},\ \bibinfo {address} {London},\ \bibinfo {year}
  {1994})\BibitemShut {NoStop}%
\bibitem [{\citenamefont {Damski}\ \emph {et~al.}(2003)\citenamefont {Damski},
  \citenamefont {Santos}, \citenamefont {Tiemann}, \citenamefont {Lewenstein},
  \citenamefont {Kotochigova}, \citenamefont {Julienne},\ and\ \citenamefont
  {Zoller}}]{zollerprop}%
  \BibitemOpen
  \bibfield  {author} {\bibinfo {author} {\bibfnamefont {B.}~\bibnamefont
  {Damski}}, \bibinfo {author} {\bibfnamefont {L.}~\bibnamefont {Santos}},
  \bibinfo {author} {\bibfnamefont {E.}~\bibnamefont {Tiemann}}, \bibinfo
  {author} {\bibfnamefont {M.}~\bibnamefont {Lewenstein}}, \bibinfo {author}
  {\bibfnamefont {S.}~\bibnamefont {Kotochigova}}, \bibinfo {author}
  {\bibfnamefont {P.}~\bibnamefont {Julienne}}, \ and\ \bibinfo {author}
  {\bibfnamefont {P.}~\bibnamefont {Zoller}},\ }\href {\doibase
  10.1103/PhysRevLett.90.110401} {\bibfield  {journal} {\bibinfo  {journal}
  {Phys. Rev. Lett.}\ }\textbf {\bibinfo {volume} {90}},\ \bibinfo {pages}
  {110401} (\bibinfo {year} {2003})}\BibitemShut {NoStop}%
\bibitem [{\citenamefont {Freericks}\ \emph {et~al.}(2010)\citenamefont
  {Freericks}, \citenamefont {Ma\ifmmode~\acute{s}\else \'{s}\fi{}ka},
  \citenamefont {Hu}, \citenamefont {Hanna}, \citenamefont {Williams},
  \citenamefont {Julienne},\ and\ \citenamefont {Lema\ifmmode~\acute{n}\else
  \'{n}\fi{}ski}}]{freericks}%
  \BibitemOpen
  \bibfield  {author} {\bibinfo {author} {\bibfnamefont {J.~K.}\ \bibnamefont
  {Freericks}}, \bibinfo {author} {\bibfnamefont {M.~M.}\ \bibnamefont
  {Ma\ifmmode~\acute{s}\else \'{s}\fi{}ka}}, \bibinfo {author} {\bibfnamefont
  {A.}~\bibnamefont {Hu}}, \bibinfo {author} {\bibfnamefont {T.~M.}\
  \bibnamefont {Hanna}}, \bibinfo {author} {\bibfnamefont {C.~J.}\ \bibnamefont
  {Williams}}, \bibinfo {author} {\bibfnamefont {P.~S.}\ \bibnamefont
  {Julienne}}, \ and\ \bibinfo {author} {\bibfnamefont {R.}~\bibnamefont
  {Lema\ifmmode~\acute{n}\else \'{n}\fi{}ski}},\ }\href {\doibase
  10.1103/PhysRevA.81.011605} {\bibfield  {journal} {\bibinfo  {journal} {Phys.
  Rev. A}\ }\textbf {\bibinfo {volume} {81}},\ \bibinfo {pages} {011605}
  (\bibinfo {year} {2010})}\BibitemShut {NoStop}%
\bibitem [{\citenamefont {Zirbel}\ \emph {et~al.}(2008)\citenamefont {Zirbel},
  \citenamefont {Ni}, \citenamefont {Ospelkaus}, \citenamefont {D'Incao},
  \citenamefont {Wieman}, \citenamefont {Ye},\ and\ \citenamefont
  {Jin}}]{zirbel}%
  \BibitemOpen
  \bibfield  {author} {\bibinfo {author} {\bibfnamefont {J.~J.}\ \bibnamefont
  {Zirbel}}, \bibinfo {author} {\bibfnamefont {K.-K.}\ \bibnamefont {Ni}},
  \bibinfo {author} {\bibfnamefont {S.}~\bibnamefont {Ospelkaus}}, \bibinfo
  {author} {\bibfnamefont {J.~P.}\ \bibnamefont {D'Incao}}, \bibinfo {author}
  {\bibfnamefont {C.~E.}\ \bibnamefont {Wieman}}, \bibinfo {author}
  {\bibfnamefont {J.}~\bibnamefont {Ye}}, \ and\ \bibinfo {author}
  {\bibfnamefont {D.~S.}\ \bibnamefont {Jin}},\ }\href {\doibase
  10.1103/PhysRevLett.100.143201} {\bibfield  {journal} {\bibinfo  {journal}
  {Phys. Rev. Lett.}\ }\textbf {\bibinfo {volume} {100}},\ \bibinfo {pages}
  {143201} (\bibinfo {year} {2008})}\BibitemShut {NoStop}%
\bibitem [{\citenamefont {Bloch}\ \emph {et~al.}(2008)\citenamefont {Bloch},
  \citenamefont {Dalibard},\ and\ \citenamefont {Zwerger}}]{latticereview}%
  \BibitemOpen
  \bibfield  {author} {\bibinfo {author} {\bibfnamefont {I.}~\bibnamefont
  {Bloch}}, \bibinfo {author} {\bibfnamefont {J.}~\bibnamefont {Dalibard}}, \
  and\ \bibinfo {author} {\bibfnamefont {W.}~\bibnamefont {Zwerger}},\ }\href
  {\doibase 10.1103/RevModPhys.80.885} {\bibfield  {journal} {\bibinfo
  {journal} {Rev. Mod. Phys.}\ }\textbf {\bibinfo {volume} {80}},\ \bibinfo
  {pages} {885} (\bibinfo {year} {2008})}\BibitemShut {NoStop}%
\bibitem [{\citenamefont {Schneider}\ \emph {et~al.}(2008)\citenamefont
  {Schneider}, \citenamefont {Hackermüller}, \citenamefont {Will},
  \citenamefont {Best}, \citenamefont {Bloch}, \citenamefont {Costi},
  \citenamefont {Helmes}, \citenamefont {Rasch},\ and\ \citenamefont
  {Rosch}}]{fermibandinsulator}%
  \BibitemOpen
  \bibfield  {author} {\bibinfo {author} {\bibfnamefont {U.}~\bibnamefont
  {Schneider}}, \bibinfo {author} {\bibfnamefont {L.}~\bibnamefont
  {Hackermüller}}, \bibinfo {author} {\bibfnamefont {S.}~\bibnamefont {Will}},
  \bibinfo {author} {\bibfnamefont {T.}~\bibnamefont {Best}}, \bibinfo {author}
  {\bibfnamefont {I.}~\bibnamefont {Bloch}}, \bibinfo {author} {\bibfnamefont
  {T.~A.}\ \bibnamefont {Costi}}, \bibinfo {author} {\bibfnamefont {R.~W.}\
  \bibnamefont {Helmes}}, \bibinfo {author} {\bibfnamefont {D.}~\bibnamefont
  {Rasch}}, \ and\ \bibinfo {author} {\bibfnamefont {A.}~\bibnamefont
  {Rosch}},\ }\href {\doibase 10.1126/science.1165449} {\bibfield  {journal}
  {\bibinfo  {journal} {Science}\ }\textbf {\bibinfo {volume} {322}},\ \bibinfo
  {pages} {1520} (\bibinfo {year} {2008})}\BibitemShut {NoStop}%
\bibitem [{\citenamefont {J\"ordens}\ \emph {et~al.}(2008)\citenamefont
  {J\"ordens}, \citenamefont {Strohmaier}, \citenamefont {G\"unter},
  \citenamefont {Moritz},\ and\ \citenamefont
  {Esslinger}}]{fermimottinsulator}%
  \BibitemOpen
  \bibfield  {author} {\bibinfo {author} {\bibfnamefont {R.}~\bibnamefont
  {J\"ordens}}, \bibinfo {author} {\bibfnamefont {N.}~\bibnamefont
  {Strohmaier}}, \bibinfo {author} {\bibfnamefont {K.}~\bibnamefont
  {G\"unter}}, \bibinfo {author} {\bibfnamefont {H.}~\bibnamefont {Moritz}}, \
  and\ \bibinfo {author} {\bibfnamefont {T.}~\bibnamefont {Esslinger}},\ }\href
  {http://dx.doi.org/10.1038/nature07244} {\bibfield  {journal} {\bibinfo
  {journal} {Nature}\ }\textbf {\bibinfo {volume} {455}},\ \bibinfo {pages}
  {204} (\bibinfo {year} {2008})}\BibitemShut {NoStop}%
\bibitem [{\citenamefont {Klempt}\ \emph {et~al.}(2008)\citenamefont {Klempt},
  \citenamefont {Henninger}, \citenamefont {Topic}, \citenamefont {Scherer},
  \citenamefont {Kattner}, \citenamefont {Tiemann}, \citenamefont {Ertmer},\
  and\ \citenamefont {Arlt}}]{fbrparameters}%
  \BibitemOpen
  \bibfield  {author} {\bibinfo {author} {\bibfnamefont {C.}~\bibnamefont
  {Klempt}}, \bibinfo {author} {\bibfnamefont {T.}~\bibnamefont {Henninger}},
  \bibinfo {author} {\bibfnamefont {O.}~\bibnamefont {Topic}}, \bibinfo
  {author} {\bibfnamefont {M.}~\bibnamefont {Scherer}}, \bibinfo {author}
  {\bibfnamefont {L.}~\bibnamefont {Kattner}}, \bibinfo {author} {\bibfnamefont
  {E.}~\bibnamefont {Tiemann}}, \bibinfo {author} {\bibfnamefont
  {W.}~\bibnamefont {Ertmer}}, \ and\ \bibinfo {author} {\bibfnamefont {J.~J.}\
  \bibnamefont {Arlt}},\ }\href {\doibase 10.1103/PhysRevA.78.061602}
  {\bibfield  {journal} {\bibinfo  {journal} {Phys. Rev. A}\ }\textbf {\bibinfo
  {volume} {78}},\ \bibinfo {pages} {061602} (\bibinfo {year}
  {2008})}\BibitemShut {NoStop}%
\bibitem [{\citenamefont {K\"ohl}\ \emph {et~al.}(2005)\citenamefont {K\"ohl},
  \citenamefont {Moritz}, \citenamefont {St\"oferle}, \citenamefont
  {G\"unter},\ and\ \citenamefont {Esslinger}}]{zurichhigherbands}%
  \BibitemOpen
  \bibfield  {author} {\bibinfo {author} {\bibfnamefont {M.}~\bibnamefont
  {K\"ohl}}, \bibinfo {author} {\bibfnamefont {H.}~\bibnamefont {Moritz}},
  \bibinfo {author} {\bibfnamefont {T.}~\bibnamefont {St\"oferle}}, \bibinfo
  {author} {\bibfnamefont {K.}~\bibnamefont {G\"unter}}, \ and\ \bibinfo
  {author} {\bibfnamefont {T.}~\bibnamefont {Esslinger}},\ }\href {\doibase
  10.1103/PhysRevLett.94.080403} {\bibfield  {journal} {\bibinfo  {journal}
  {Phys. Rev. Lett.}\ }\textbf {\bibinfo {volume} {94}},\ \bibinfo {pages}
  {080403} (\bibinfo {year} {2005})}\BibitemShut {NoStop}%
\bibitem [{\citenamefont {DeMarco}\ \emph {et~al.}(2005)\citenamefont
  {DeMarco}, \citenamefont {Lannert}, \citenamefont {Vishveshwara},\ and\
  \citenamefont {Wei}}]{rbcalc}%
  \BibitemOpen
  \bibfield  {author} {\bibinfo {author} {\bibfnamefont {B.}~\bibnamefont
  {DeMarco}}, \bibinfo {author} {\bibfnamefont {C.}~\bibnamefont {Lannert}},
  \bibinfo {author} {\bibfnamefont {S.}~\bibnamefont {Vishveshwara}}, \ and\
  \bibinfo {author} {\bibfnamefont {T.-C.}\ \bibnamefont {Wei}},\ }\href
  {\doibase 10.1103/PhysRevA.71.063601} {\bibfield  {journal} {\bibinfo
  {journal} {Phys. Rev. A}\ }\textbf {\bibinfo {volume} {71}},\ \bibinfo
  {pages} {063601} (\bibinfo {year} {2005})}\BibitemShut {NoStop}%
\bibitem [{som()}]{som}%
  \BibitemOpen
  \href@noop {} {\bibinfo  {journal} {See the Supplementary Materials}\
  }\BibitemShut {NoStop}%
\bibitem [{\citenamefont {Sugawa}\ \emph {et~al.}(2011)\citenamefont {Sugawa},
  \citenamefont {Inaba}, \citenamefont {Taie}, \citenamefont {Yamazaki},
  \citenamefont {Yamashita},\ and\ \citenamefont
  {Takahashi}}]{dualybinsulator}%
  \BibitemOpen
\bibfield  {journal} {  }\bibfield  {author} {\bibinfo {author} {\bibfnamefont
  {S.}~\bibnamefont {Sugawa}}, \bibinfo {author} {\bibfnamefont
  {K.}~\bibnamefont {Inaba}}, \bibinfo {author} {\bibfnamefont
  {S.}~\bibnamefont {Taie}}, \bibinfo {author} {\bibfnamefont {R.}~\bibnamefont
  {Yamazaki}}, \bibinfo {author} {\bibfnamefont {M.}~\bibnamefont {Yamashita}},
  \ and\ \bibinfo {author} {\bibfnamefont {Y.}~\bibnamefont {Takahashi}},\
  }\href {http://dx.doi.org/10.1038/nphys2028} {\bibfield  {journal} {\bibinfo
  {journal} {Nat. Phys.}\ }\textbf {\bibinfo {volume} {7}},\ \bibinfo {pages}
  {642} (\bibinfo {year} {2011})}\BibitemShut {NoStop}%
\bibitem [{\citenamefont {Neyenhuis}\ \emph {et~al.}(2012)\citenamefont
  {Neyenhuis}, \citenamefont {Yan}, \citenamefont {Moses}, \citenamefont
  {Covey}, \citenamefont {Chotia}, \citenamefont {Petrov}, \citenamefont
  {Kotochigova}, \citenamefont {Ye},\ and\ \citenamefont {Jin}}]{neyenhuis}%
  \BibitemOpen
  \bibfield  {author} {\bibinfo {author} {\bibfnamefont {B.}~\bibnamefont
  {Neyenhuis}}, \bibinfo {author} {\bibfnamefont {B.}~\bibnamefont {Yan}},
  \bibinfo {author} {\bibfnamefont {S.~A.}\ \bibnamefont {Moses}}, \bibinfo
  {author} {\bibfnamefont {J.~P.}\ \bibnamefont {Covey}}, \bibinfo {author}
  {\bibfnamefont {A.}~\bibnamefont {Chotia}}, \bibinfo {author} {\bibfnamefont
  {A.}~\bibnamefont {Petrov}}, \bibinfo {author} {\bibfnamefont
  {S.}~\bibnamefont {Kotochigova}}, \bibinfo {author} {\bibfnamefont
  {J.}~\bibnamefont {Ye}}, \ and\ \bibinfo {author} {\bibfnamefont {D.~S.}\
  \bibnamefont {Jin}},\ }\href {\doibase 10.1103/PhysRevLett.109.230403}
  {\bibfield  {journal} {\bibinfo  {journal} {Phys. Rev. Lett.}\ }\textbf
  {\bibinfo {volume} {109}},\ \bibinfo {pages} {230403} (\bibinfo {year}
  {2012})}\BibitemShut {NoStop}%
\bibitem [{\citenamefont {Budker}\ \emph {et~al.}(2008)\citenamefont {Budker},
  \citenamefont {Kimball},\ and\ \citenamefont {DeMille}}]{fillinguniform}%
  \BibitemOpen
  \bibfield  {author} {\bibinfo {author} {\bibfnamefont {D.}~\bibnamefont
  {Budker}}, \bibinfo {author} {\bibfnamefont {D.}~\bibnamefont {Kimball}}, \
  and\ \bibinfo {author} {\bibfnamefont {D.}~\bibnamefont {DeMille}},\
  }\href@noop {} {\emph {\bibinfo {title} {Atomic Physics: An Exploration
  through Problems and Solutions, Second Edition}}}\ (\bibinfo  {publisher}
  {Oxford University Press},\ \bibinfo {address} {Oxford},\ \bibinfo {year}
  {2008})\BibitemShut {NoStop}%
\bibitem [{\citenamefont {Bakr}\ \emph {et~al.}(2009)\citenamefont {Bakr},
  \citenamefont {Gillen}, \citenamefont {Peng}, \citenamefont {F\"olling},\
  and\ \citenamefont {Greiner}}]{greinermicroscope}%
  \BibitemOpen
  \bibfield  {author} {\bibinfo {author} {\bibfnamefont {W.~S.}\ \bibnamefont
  {Bakr}}, \bibinfo {author} {\bibfnamefont {J.~I.}\ \bibnamefont {Gillen}},
  \bibinfo {author} {\bibfnamefont {A.}~\bibnamefont {Peng}}, \bibinfo {author}
  {\bibfnamefont {S.}~\bibnamefont {F\"olling}}, \ and\ \bibinfo {author}
  {\bibfnamefont {M.}~\bibnamefont {Greiner}},\ }\href
  {http://dx.doi.org/10.1038/nature08482} {\bibfield  {journal} {\bibinfo
  {journal} {Nature}\ }\textbf {\bibinfo {volume} {462}},\ \bibinfo {pages}
  {74} (\bibinfo {year} {2009})}\BibitemShut {NoStop}%
\bibitem [{\citenamefont {Sherson}\ \emph {et~al.}(2010)\citenamefont
  {Sherson}, \citenamefont {Weitenberg}, \citenamefont {Endres}, \citenamefont
  {Cheneau}, \citenamefont {Bloch},\ and\ \citenamefont
  {Kuhr}}]{blochmicroscope}%
  \BibitemOpen
  \bibfield  {author} {\bibinfo {author} {\bibfnamefont {J.~F.}\ \bibnamefont
  {Sherson}}, \bibinfo {author} {\bibfnamefont {C.}~\bibnamefont {Weitenberg}},
  \bibinfo {author} {\bibfnamefont {M.}~\bibnamefont {Endres}}, \bibinfo
  {author} {\bibfnamefont {M.}~\bibnamefont {Cheneau}}, \bibinfo {author}
  {\bibfnamefont {I.}~\bibnamefont {Bloch}}, \ and\ \bibinfo {author}
  {\bibfnamefont {S.}~\bibnamefont {Kuhr}},\ }\href
  {http://dx.doi.org/10.1038/nature09378} {\bibfield  {journal} {\bibinfo
  {journal} {Nature}\ }\textbf {\bibinfo {volume} {467}},\ \bibinfo {pages}
  {68} (\bibinfo {year} {2010})}\BibitemShut {NoStop}%
\bibitem [{\citenamefont {Gemelke}\ \emph {et~al.}(2009)\citenamefont
  {Gemelke}, \citenamefont {Zhang}, \citenamefont {Hung},\ and\ \citenamefont
  {Chin}}]{chinmott}%
  \BibitemOpen
  \bibfield  {author} {\bibinfo {author} {\bibfnamefont {N.}~\bibnamefont
  {Gemelke}}, \bibinfo {author} {\bibfnamefont {X.}~\bibnamefont {Zhang}},
  \bibinfo {author} {\bibfnamefont {C.-L.}\ \bibnamefont {Hung}}, \ and\
  \bibinfo {author} {\bibfnamefont {C.}~\bibnamefont {Chin}},\ }\href
  {http://dx.doi.org/10.1038/nature08244} {\bibfield  {journal} {\bibinfo
  {journal} {Nature}\ }\textbf {\bibinfo {volume} {460}},\ \bibinfo {pages}
  {995} (\bibinfo {year} {2009})}\BibitemShut {NoStop}%
\end{thebibliography}%

\section*{Supplementary Materials}

\subsection*{Rb MI calculation}

For a perfect Rb MI, we calculate the distribution at zero temperature and without tunneling, based on Ref.~\cite{rbcalc}.  We numerically find the relationship between the chemical potential $\mu_0$ and particle number $N$.  The local $\mu$ at lattice site $(i,j,k)$ is $\mu(i,j,k)=\mu_0-V(i,j,k)$, where $V$ is the harmonic confinement.  In the zero tunneling approximation, the occupancy on site $(i,j,k)$, $n$, satisfies $(n-1) < \frac{\mu(i,j,k)}{U} \le n$. The green staircase in Fig.~3A displays the peak $T=0$ occupancy. The green shaded areas indicate experimental uncertainty of the harmonic trap frequency $\omega_r = 2 \pi \times 38(2)$ Hz and aspect ratio $A=6.4(1)$.  To make a closer comparison to the experiment, we sum the number of atoms along the $z$ direction (following the experimental geometry where the probe beam integrates along $z$), and convolve the resulting 2D distribution with a Gaussian filter with rms width 4.5(5) lattice sites to simulate the effect of finite imaging resolution.  We also account for pixelation by mapping arrays of $6 \times 6$ lattice sites onto single pixels.  This gives us a convolved, pixelated 2D distribution that we then fit with a 2D TF surface to extract $f_{\text{TF}}$, shown as the orange shaded band in Fig.~3A. The width of the band again accounts for the uncertainties in the trap.  

\subsection*{Molecule production and detection}

Similar to previous work \cite{kangkuen,marcio, amodsenprl}, we create weakly bound Feshbach molecules by magneto-association, in which the magnetic field is swept from above the resonance to below the resonance.  In the experiments reported here, the sweep takes 5 ms, starts at 563 G, and ends at 545.6 G.  We then perform STIRAP to transfer the Feshbach molecules to the ro-vibrational ground state. The two STIRAP lasers, at 968 nm and 689 nm, are frequency stabilized to a common high-finesse optical cavity.  After STIRAP, we remove the unpaired K atoms with a pulse of resonant light and we remove the Rb atoms with a series of microwave adiabatic rapid passages (ARPs) to transfer the atoms to the $\ket{2,2}$ state followed with pulses of resonant light.  We find that these removals are required in order for the molecules to have a long lifetime in the lattice.  To detect the ground-state molecules, we reverse the STIRAP process to transfer the ground-state molecules back to the Feshbach molecule state, sweep the magnetic field back to 563 G in 1 ms to dissociate the Feshbach molecules, and then image the resulting K atoms.  The molecules can also be detected by measuring the resulting Rb atoms, and the numbers agree within the experimental uncertainty.

\end{document}